\documentclass[11pt]{article}

\usepackage[T1]{fontenc}
\usepackage{palatino}

\usepackage[margin=2.5cm]{geometry}
\usepackage{setspace}
\usepackage{graphicx}
\usepackage{color}
\usepackage[breaklinks]{hyperref}
\usepackage{nowidow}
\usepackage{tabularx}
\usepackage{dcolumn}
\usepackage{makecell}
\usepackage[document]{ragged2e}

\usepackage{caption}
\usepackage[labelfont=bf]{caption}

\usepackage{mwe}
\usepackage{floatrow}
\newfloatcommand{capbtabbox}{table}[][\FBwidth]

\usepackage{orcidlink}
\usepackage{authblk}

\usepackage{amsmath}
\usepackage[group-separator={,}]{siunitx}

\PassOptionsToPackage{natbib=true}{biblatex}
\RequirePackage[authordate,giveninits,backend=biber]{biblatex-chicago}
\title{\textbf{Amenity complexity and urban locations \\ of socio-economic mixing}}

\author[1,2,3,*]{Sándor Juhász \orcidlink{0000-0003-3124-8597}}
\author[2]{Gergő Pintér \orcidlink{0000-0003-4731-3816}}
\author[3]{Ádám Kovács \orcidlink{0000-0001-6503-3047}}
\author[3]{Endre Borza \orcidlink{0000-0002-8804-4520}}
\author[2]{\\Gergely Mónus \orcidlink{0000-0002-4768-736X}}
\author[2,3]{László Lőrincz \orcidlink{0000-0003-4528-0918}}
\author[2,3]{Balázs Lengyel \orcidlink{0000-0001-5196-5599}}
\affil[1]{Complexity Science Hub Vienna, Vienna, Austria}
\affil[2]{NETI Lab, Corvinus University of Budapest, Budapest, Hungary}
\affil[3]{ANET Lab, ELKH Center for Economic and Regional Sciences, Budapest, Hungary}
\affil[*]{Corresponding author: \href{mailto:juhasz@csh.ac.at}{juhasz@csh.ac.at}}

\date{}

\begin{document}

\maketitle

\onehalfspacing

\begin{abstract}
\noindent
\justify
Cities host diverse people and their mixing is the engine of prosperity. In turn, segregation and inequalities are common features of most cities and locations that enable the meeting of people with different socio-economic status are key for urban inclusion. In this study, we adopt the concept of economic complexity to quantify the sophistication of amenity supply at urban locations. We propose that neighborhood complexity and amenity complexity are connected to the ability of locations to attract diverse visitors from various socio-economic backgrounds across the city.
We construct the measures of amenity complexity based on the local portfolio of diverse and non-ubiquitous amenities in Budapest, Hungary. Socio-economic mixing at visited third places is investigated by tracing the daily mobility of individuals and by characterizing their status by the real-estate price of their home locations. Results suggest that measures of ubiquity and diversity of amenities do not, but neighborhood complexity and amenity complexity are correlated with the urban centrality of locations. Urban centrality is a strong predictor of socio-economic mixing, but both neighborhood complexity and amenity complexity add further explanatory power to our models. Our work combines urban mobility data with economic complexity thinking to show that the diversity of non-ubiquitous amenities, central locations, and the potentials for socio-economic mixing are interrelated.
\end{abstract}

\section{Introduction}
\justify

% diversity is key
Diversity is the key ingredient of successful and resilient cities \citep{jacobs1961death}. The spatially concentrated interaction of people from various social and economic background create environments that foster creativity \citep{florida2004creative}, support inclusion \citep{bentonshort2013citiesnature} and in general, make cities vivid and prosperous \citep{glaeser2012triumph}.
At the same time, cities show high levels of segregation such that individuals from different socio-economic background are separated from each other in the urban space \citep{musterd2020handbook}. This phenomenon limits social mobility for many \citep{mayer1989science} and induced inequalities can expose segregated groups to health or climate crises \citep{torrats2021using, loughran2022unequal} and can imply radicalization and populism \citep{abadie2006poverty, engler2021threat}.

% urban mobility data -- results from a new frontier
Recent studies leverage GPS mobility data to study socio-economic segregation and mixing patterns in visited urban locations \citep{cagney2020mobilityreview}. This growing literature frequently reports that people in cities visit and interact with locations that are similar to their residential neighborhood in terms of income, education, ethnicity or other socio-economic features \citep{wang2018pnas, dong2020epj, bokanyi2021scirep, hilman2022socioeconomic}. However, the places, services or amenities that individuals visit in the city exhibit different levels of experienced segregation, as some locations mix different socio-economic groups while others do not \citep{athey2021estimating, moro2021naturecomm}.

% WHAT kind of locations mix people?
In this study we characterize urban locations that foster socio-economic mixing and lower experienced segregation by attracting people from diverse strata. To do so, we emphasize two aspects of urban locations that can influence observed socio-economic mixing: their amenity portfolio and geographical centrality in the city.

The type of amenities available at a location determine its purpose and function and therefore is related to experienced segregation. \textcite{noyman2019epb} illustrates through individual GPS trajectories that urban locations offering entertainment amenities, services or natural water features are visited by a more diverse set of people. \textcite{athey2021estimating} describes that individuals can experience relatively low experienced segregation at outdoor places like parks, sports fields and playgrounds, or at commercial establishments such as restaurants, bars and retail stores. They find that places of entertainment, like theaters and accommodations, like hotels are the least segregated urban locations. \textcite{moro2021naturecomm} shows that the category of places is a strong predictor for experienced income segregation and unique places in cities, such as arts venues, museums or airports tend to be highly integrative, while places that primarily serve local communities, such as grocery stores or places of worship are generally more segregated by income. Yet, urban locations can be hardly described by single amenity types; instead, they typically host more types of amenities. Despite previous empirical efforts, systematic examination on how the mixture of amenities at specific urban locations contribute to social mixing is still missing from the literature.

% WHERE are urban locations that attract diverse visitors located?
Specialized amenities that serve the specific needs of the wider public and therefore can attract people from diverse neighborhoods tend to situate in the center of cities. The central place theory originally developed for the inter-urban scale by \textcite{christaller1933cpt} and \textcite{losch1954cpt} explains the hierarchy of cities and towns through their size and the range of functions that they provide and has been used to study the functions of locations within cities too (see for example \citep{wang2021polycentric}). Higher-order centers attract population from a larger area, because they not only share most of the functions of lower order centers, but also host some more specialized functions too. Building on the central place theory, \textcite{zhong2015centrality} combines density, the number of people attracted to locations and diversity, the range of activities that they engage with at these locations in a single centrality measure to identify urban centers in Singapore and illustrate their evolution over time. \textcite{noyman2019epb} shows that urban locations with higher centrality in urban road networks attract more diverse visitors. On the contrary, \textcite{moro2021naturecomm} presents that urban locations with higher average travel distance to them tend to be less segregated than locations that are highly accessible. While most of the studies highlight that accessible, central locations attract more diverse visitors, yet, the nature of the available amenity mix might be related to the position of locations, which has not been focused on so far.

% question and aim
Here, we aim to extend the above literature by investigating how the available amenities and central position of urban locations are related to experienced segregation or, put it differently, to the mixing of people from diverse socio-economic strata. A new contribution is the application of the economic complexity framework to urban amenities \citep{hidalgo2021review} to quantify the sophistication of local amenity supply. We argue that more complex neighborhoods and amenities attract visitors of diverse socio-economic status from across the city.

% complexity, building blocks and urban amenities
The concept of economic complexity is originally developed by \textcite{hidalgohausmann2009pnas} who defined complexity of economies by the diversity of their non-ubiquitous products and services. Economic complexity is indicative of countries economic growth, income level, emissions and inequalities \citep{hidalgo2021review}. By now, the concept is applied to different data sources such as patents, occupations or scientific publications and to diverse spatial scales from countries to cities \citep{balland2022review, lucas2023ecibrazil}.
Here we adopt the measurement technique to uncover the complexity of neighborhoods and amenities. We propose that a neighborhood has a complex amenity mix in case it offers diverse set of amenities of those types that other locations are not specialized in. On the contrary, complex amenities are those that only few neighborhoods are specialized into and are co-located with diverse sets of similarly non-ubiquitous amenities. 
Unlike in the original framework of economic complexity that captures the knowledge and capabilities required to achieve economic outputs \citep{hidalgohausmann2009pnas}, our approach does not address the productive and operative knowledge that a given location has accumulated \citep{hidalgo2021review}. Instead, we measure the sophistication of local amenity supply that can serve a wide range of unique needs.

% expectations
The rational to apply neighborhood and amenity complexity to understand mixing of people is based on two reasons. First, diverse amenity mixes in neighborhoods can attract people with diverse demands. Second, locations with non-ubiquitous amenities can attract people from diverse neighborhoods, as the particular service is hard to find elsewhere. Consequently, complex amenities combining diversity and non-ubiquity are also expected to attract diverse visitors.
Therefore, our hypothesis is that the diverse mix of non-ubiquitous amenities can create an inclusive, multipurpose neighborhood that is most likely to be attractive for a wide-variety of people. While the contribution of amenity mix to the socio-economic diversity of visitors at urban locations has rarely been unveiled, diverse amenities are argued to concentrate in and attract people to central places of cities \citep{zhong2015centrality}. To better understand the connection between neighborhood and amenity complexity, urban centrality and socio-economic mixing, we test their correlation with the socio-economic diversity of visitors.

We test this argument in Budapest, the capital of Hungary by combining point of interests (POI) data collected from the Google Places API and individual mobility trajectories collected by a GPS aggregator company. Building on the work of \textcite{hidalgo2020amenitymix}, we construct the indicators of neighborhood and amenity complexity by utilizing the geographic distribution of POIs in neighborhoods. We identify home, work and third place visits in daily mobility trajectories for 24 months by clustering the geolocated pings of devices in geographical space and over time \citep{oldenburg1999thirdplaces}. We combine the information of predicted home locations with real estate prices at the census tract level. This allows us to investigate third place visits and to infer the socio-economic diversity of visitors in each urban neighborhood and in each actual amenity. 

% contributions
Our results illustrate that, in the monocentric city of Budapest, specialization and diversity of amenities do not, but amenity complexity is correlated with urban centrality. We find that socio-economic mixing increases as neighborhood complexity grows, and that amenity complexity is also associated with lower levels of experienced segregation. These suggest that the combination of mobility data with economic complexity thinking can provide new insights to the research of urban segregation and mixing.

\section{Tracing mobility inside cities}
\justify

Urban mobility of individuals are studied by using raw GPS data from a data aggregator company. We can trace the daily mobility of 5.2 million devices in Hungary over 24 months (between 2019 June and 2021 May). We initially filter this data to focus on devices that appear inside Budapest and have at least 20 GPS pings in total after discarding pings which indicate unreasonably high speeds of device mobility. Detailed description on the mobility data preparation process can be found in section 1 of the Supplementary information.

\begin{figure}[!ht]
\includegraphics[width=0.92\textwidth]{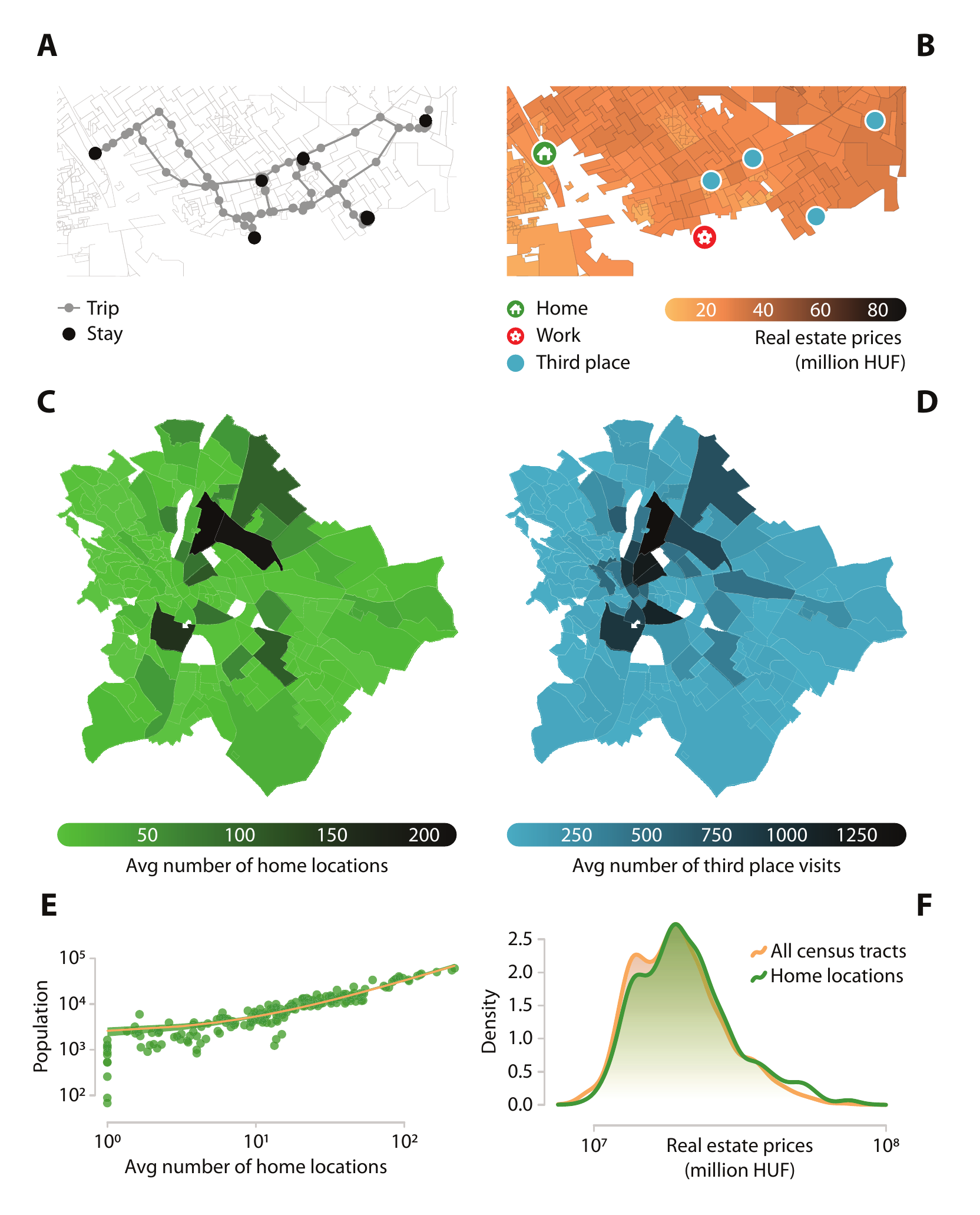}  
\caption{Identifying home locations and third places visits from daily mobility trajectories.
\textbf{(A)} Example trajectory to illustrate the stop detection process.
\textbf{(B)} Identified stops and predicted home, work, and visited third places.
\textbf{(C)} Average number of home location and 
\textbf{(D)} average number of third place visits over 24 months by urban neighborhoods of Budapest.
\textbf{(E)} The relationship between average number of home locations over 24 months and population of urban neighborhoods in Budapest.
\textbf{(F)} Real estate prices at census tracts of identified home locations and across all census tracts of Budapest.}
\label{fig:fig1}
\end{figure}

We process raw trajectories of individuals by applying the Infostop algorithm \citep{aslak2020infostop}. It enables us to detect the stationary points of individual movements and cluster GPS pings around stop locations. Figure~\ref{fig:fig1}A-B illustrates the raw data and the outcome of stop detection through an example device. The algorithm gives each stop a label indicating a place that can reoccur along the trajectory of the device. We focus on devices with at least 2 distinct places and 10 stops in a month inside Budapest. Using the monthly recurrence of stops and places by each device, we label places as home, work or third place visits in two steps. 

First, we categorize each visited place as potential home or work based on the part of the day it is visited, the duration of visits and their reappearance in the daily trajectory. The potential home is where the device spends the most time between 8:00pm and 8:00am on weekdays or at any time during the weekend, and the cumulative time spent at the place exceeds 8 hours per week. Places where devices spend the most time between 9:00am and 5:00pm on weekdays (at least 3 hours a week) are considered as potential workplaces.

Second, we time-aggregate device trajectories to monthly visitation patterns. Thus, we identify home and work of a device in a month by the mean coordinate pairs of weekly potential home and work places, but only in case a device stops at the place at least 10 times over a month and the standard deviation of both latitude and longitude coordinates are smaller than 0.001 (about 100 meters in Budapest) over the respective month. We categorize every other visited place as a third place, in case it is labeled by the stop detection algorithm as a unique place, but it is not the home or the work place of the device in the respective month. Figure~\ref{fig:fig1}C presents the average number of devices with identified home location (and at least one visited third place) and Figure~\ref{fig:fig1}D illustrates the average number of third place visits over the 24-month period aggregated to the level of urban neighborhoods.

Home locations and third places are joined to other data sources with Uber's Hexagonal Hierarchical Spatial Index (H3) \citep{h3}. The applied indexes of size 10 H3 hexagons refer to an average 15.000 m\textsuperscript{2} area, which is close to the buffer area of a point with a 70 meter radius. We connect all the identified home locations and third places to hexagons and split each neighborhood or census tract level polygons to the same hexagon size for efficient combination.

To infer the socio-economic status of the followed devices, we join home locations to census tract level real estate prices. In Hungary, information on income is not part of the census data collection. We rely on residential real estate sales contracts from 2013-2019 collected by the Hungarian Central Statistical Office and predict real estate prices to each census tract of Budapest. Section 2 in the Supplementary information introduces the prediction process in detail. Figure~\ref{fig:fig1}F presents that real estate prices at the identified home locations and across all census tracts are closely align.

\section{Measuring amenity complexity}
\justify

To describe the attractiveness of urban locations, we construct the measures of amenity complexity. These indicators are based on the spatial distribution of amenities, which is studied through point of interest (POI) data from the Google Places API. Besides its limitations in terms of timescale and POI categorization, it is one of the world's most popular mapping service supporting applications worldwide and helping millions of individuals on a daily basis to find the location of businesses. This makes Google data attractive to study the spatial organization of amenities inside cities \citep{hidalgo2020amenitymix, kaufmann2022amenityscaling, heroy2022walkordrive}.

We collected GPS coordinates and amenity category for all the POIs around the city of Budapest in early 2022. The resulted data set contains 63.601 POIs in 78 different amenity categories. We removed the frequently appearing and ambiguous categories of ATM (1.054 POIs) and Parking (729 POIs) and filter out the category Casino with less than 2 POIs in Budapest. We use this data to illustrate the amenity profile of the 207 urban neighborhoods of Budapest \citep{kshregionalatlas}. Neighborhoods are in between districts and census tracts in the spatial hierarchy, which makes them a suitable scale for our analysis \citep{natera2020walkability}. They have an average population of 10.000 people (standard deviation around 10.000), have an average area of 2.5 km\textsuperscript{2} (standard deviation around 3.9) and on average they consist of 41 lower level census tracts (standard deviation around 50). Further description about the neighborhoods of Budapest can be found in section 3 of the Supplementary information.

Every neighborhood in Budapest with at least 2 amenity categories that have minimum 2 POIs are considered in the analysis. Alternative specifications and their influence on amenity complexity measurement can be found in section 4 of the Supplementary Information. Figure~\ref{fig:fig2}A presents the resulted 75 amenity categories and the number of POIs across the focal 200 neighborhoods. The most frequent categories are convenient store (5.989 observations), beauty salon (4.461 observations) and restaurant (3.727 observations), while we observe less than 10 amusement parks, bowling alley and city hall. Figure~\ref{fig:fig2}B illustrates the unequal spatial distribution of POIs on the map of neighborhoods in Budapest.

\begin{figure}[!ht]
\includegraphics[width=0.925\textwidth]{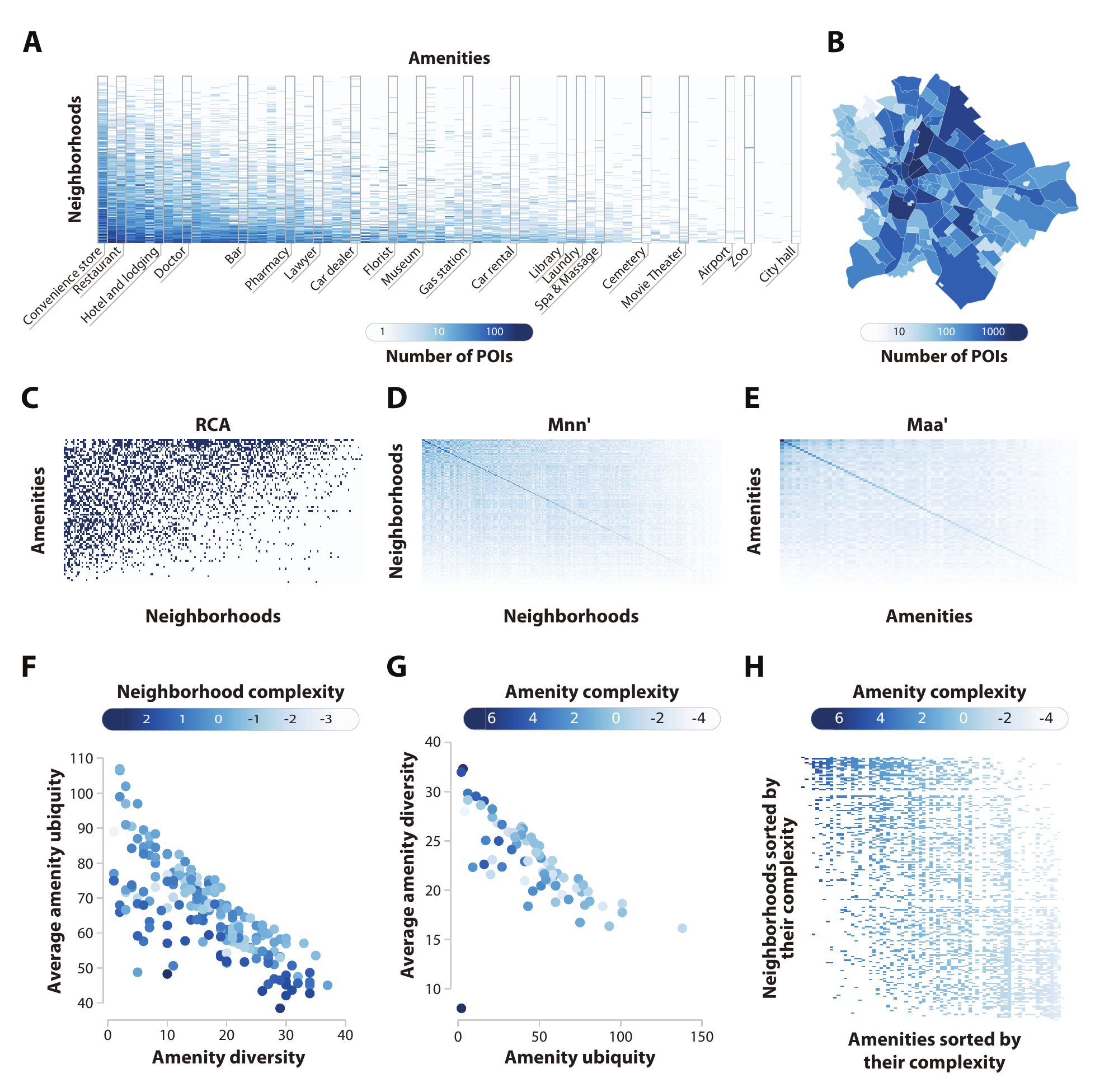}  
\caption{Constructing the measures of neighborhood and amenity complexity.
\textbf{(A)} Distribution of point of interests (POIs) across neighborhoods and amenity categories.
\textbf{(B)} Map of urban neighborhoods colored by the number of observed POIs.
\textbf{(C)} Revealed comparative advantage (RCA) values transformed to a binary specialization matrix (M).
\textbf{(D)} Similarity matrix of neighborhoods based on their specialization in amenity categories. This matrix is used to measure neighborhood complexity.
\textbf{(E)} Similarity matrix of amenities based on their specialization in neighborhoods. This matrix is used to measure amenity complexity.
\textbf{(F)} Relationship between amenity diversity and average amenity ubiquity in neighborhoods. Dots (neighborhoods) are colored by their neighborhood complexity value.
\textbf{(G)} Relationship between ubiquity and average diversity of amenities. Dots (amenity categories) are colored by their amenity complexity value.
\textbf{(H)} Neighborhoods with higher complexity value are specialized in amenity categories that have a higher amenity complexity value. Each cell in the matrix represents a neighborhood specialized in an amenity category and cells are colored by amenity complexity.}
\label{fig:fig2}
\end{figure}

To describe the relative importance of amenity categories and illustrate the differences between the amenity structure of neighborhoods, we adopt the economic complexity index (ECI) and the product complexity index (PCI) \citep{hidalgohausmann2009pnas}. The ECI is successfully used to describe the economic development of countries and regions \citep{hidalgo2021review} and its approach is adoptable to amenities and urban neighborhoods. We measure neighborhood complexity and amenity complexity the following way. We normalize the matrix of Figure~\ref{fig:fig2}A to make comparisons appropriate between neighborhoods and amenity categories and compute the revealed comparative advantage (RCA) of neighborhoods in amenity categories by the following standard equation (also known as the Balassa index):

\begin{equation}
\label{eq:eq1}
    RCA_{n,a}=(P_{n,a}/P_{a})/(P_{n}/P)
\end{equation}

\noindent
where $P_{n,a}$ is the number of POIs in neighborhood $n$ in amenity category $a$ and missing indices indicate summed variables such as $P_{a}=\sum_{a}P_{n,a}$. \textit{RCA $>=$ 1} suggests that neighborhood $n$ is specialized in amenity category $a$. In other words, an amenity category is overrepresented in a neighborhood in case its \textit{RCA} value is above or equal to 1. We use the $RCA$ values to create a binary specialization matrix $M_{n,a}$ the following way:

\begin{equation}
\label{eq:eq2}
    M_{n,a} = 
\left\{
\begin{array}{ccl}
    1 & \textrm{if} & RCA_{n,a} >= 1 \\
    0 & \textrm{if} & RCA_{n,a} < 1
\end{array}
\right.
\end{equation}

\noindent
Figure~\ref{fig:fig2}C illustrates the resulted binary $RCA$ matrix of neighborhoods and amenity categories in Budapest. Sum of rows in this matrix presents the number of amenity categories a neighborhood has comparative advantage in (amenity diversity) and the column sums give the number of neighborhoods where an amenity category is overrepresented (amenity ubiquity).

\begin{equation}
\label{eq:eq3}
    \textrm{Amenity diversity} = M_{n} = \sum_{a}M_{n,a}
\end{equation}

\begin{equation}
\label{eq:eq4}
    \textrm{Amenity ubiquity} = M_{a} = \sum_{n}M_{n,a}
\end{equation}

\noindent
In geographic matrices like $M$ the average ubiquity of the activities present in a location tends to correlate negatively with the diversity of activities in a location. This is the result of the matrix property known as nestedness and this feature is utilized to explain that more complex activities are only available at a handful of locations with a diverse portfolio of activities \citep{hidalgohausmann2009pnas, balland2020nhb}.

\begin{equation}
\label{eq:eq5}
    \textrm{Neighborhood complexity} = K_{n} = \frac{1}{M_{n}}\sum_{a}M_{n,a}K_{a}
\end{equation}

\begin{equation}
\label{eq:eq6}
    \textrm{Amenity complexity} = K_{a} = \frac{1}{M_{a}}\sum_{n}M_{n,a}K_{n}
\end{equation}

The economic complexity index (ECI) that describes the production structure of economies and the product complexity index (PCI) that describe the complexity of products were originally defined through the iterative, self-referential algorithm of the 'method of reflection' \citep{hidalgohausmann2009pnas}. The algorithm calculates the above explained diversity and ubiquity vectors and then recursively uses the information in one equation to correct the other (see \ref{eq:eq5}) and (\ref{eq:eq6}). Later it was presented that the method of reflection is equivalent to finding the eigenvectors of the similarity matrix $M_{nn'}$ and $M_{aa'}$ \citep{mealy2019interpreting, hidalgo2021review}. In our case $M_{nn'}$ is defined from the original binary neighborhood-amenity matrix $M$ as $M_{nn'} = M^T * M$. The neighborhood-neighborhood similarity matrix used to construct our \textit{neighborhood complexity} measure is visualized by Figure~\ref{fig:fig2}D. Neighborhood complexity is analogous to economic complexity in terms of measurement and it captures the amenity complexity of neighborhoods. To measure the complexity of amenity categories based on their geographic distribution across neighborhoods, we create an amenity-amenity similarity matrix as $M_{aa'} = M * M^T$, visualized by Figure~\ref{fig:fig2}E. The network representation and the clustered version of this matrix can be found in Section 5 in the Supplementary information. Our \textit{amenity complexity} measure is constructed in a similar way to the product complexity index. As discussed in the Introduction, neighborhood complexity and amenity complexity are interpreted as measures of the sophistication of the local amenity supply that can serve a wide range of unique needs.

Applying the most common approach to measure complexity from geographical matrices, we take the second eigenvector of $M_{nn'}$, which is the leading correction to the equilibrium distribution and is the vector that is the best at dividing neighborhoods into groups based on the amenities that are present in them. Similarly, we take the second eigenvector of $M_{aa'}$ to get the amenity complexity values of amenity categories. This process to measure complexity is similar to dimension reduction techniques (singular value decomposition) that provide ways to explain the structure of matrices (for an overview, see \citep{hidalgo2021review}).

Figure~\ref{fig:fig2}F illustrates the relationship between amenity diversity and average amenity ubiquity of neighborhoods. Each point represents a neighborhood and is colored by the derived neighborhood complexity values. Besides the expected negative correlation between amenity diversity and average amenity ubiquity \citep{hidalgo2021review}, neighborhood complexity and the diversity of amenities at these locations shows remarkable variance. Figure~\ref{fig:fig2}G presents the relationship between the ubiquity of amenity categories and their average diversity. Each point stands for an amenity category and is colored by the derived amenity complexity values. Overall, we observe that more complex amenity categories are non-ubiquitous and on average appear in more diverse areas. However, the figure indicates a clear outlier (bottom left corner), zoo, which is very non-ubiquitous and at the same time appears in less diverse neighborhoods. Figure~\ref{fig:fig2}H visualizes the mechanical relationship between amenity complexity and neighborhood complexity. The figure makes it clear that complex neighborhoods have complex amenities. These patterns are in line with the ones revealed by \textcite{mealy2019interpreting} for countries and exported products. Section 6 in the Supplementary information presents the ranking of neighborhoods and amenities in Budapest by their neighborhood complexity and amenity complexity values.

\begin{figure}[!ht]
\includegraphics[width=0.915\textwidth]{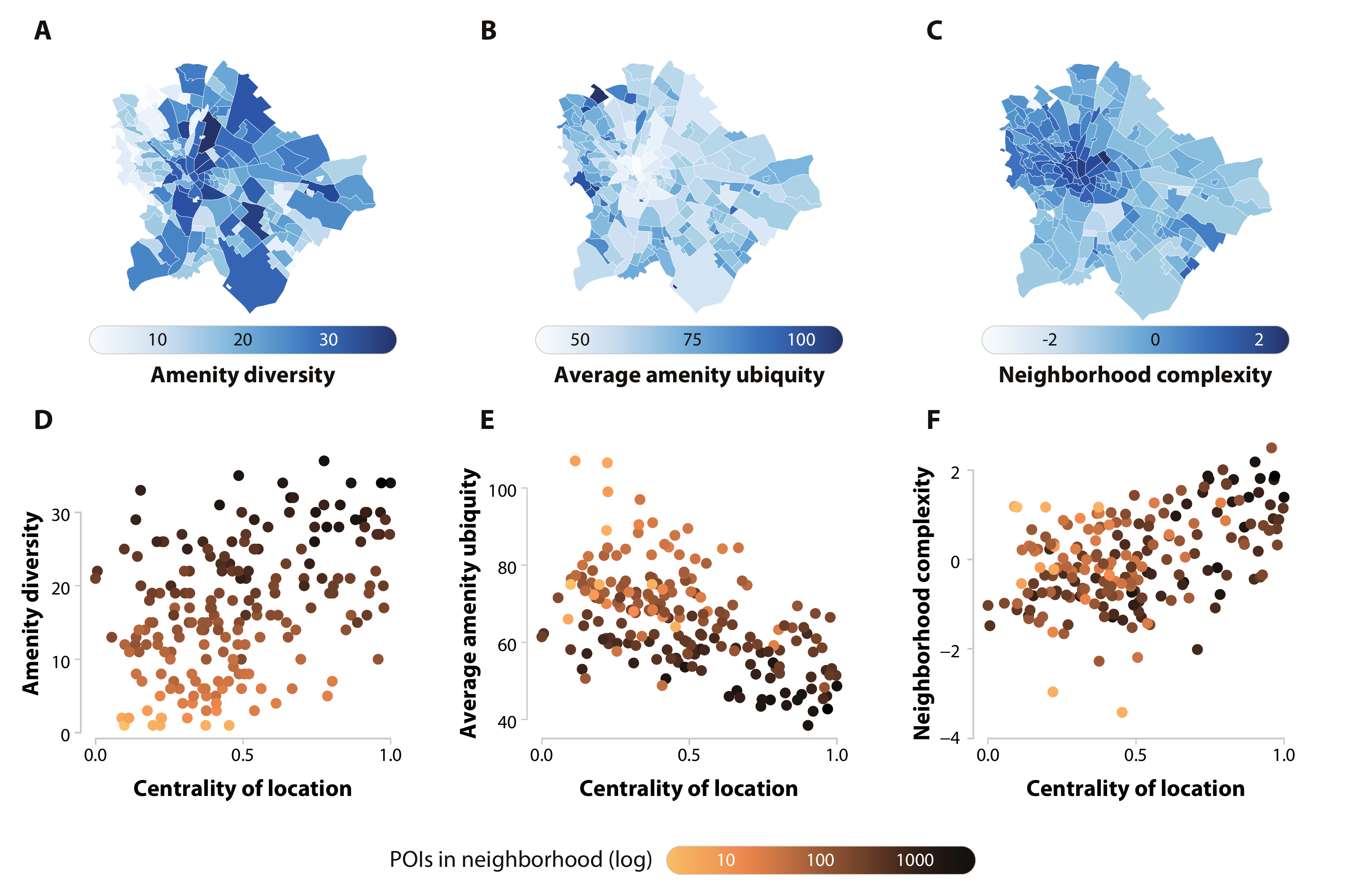}  
\caption{Components of neighborhood complexity and their relationship with geographical centrality in Budapest.
\textbf{(A)} Map of Budapest colored by the amenity diversity, \textbf{(B)} by the average amenity ubiquity, \textbf{(C)} by the amenity complexity of neighborhoods.
\textbf{(D)} Relationship between the geographical centrality and neighborhoods' amenity diversity, \textbf{(E)}  average amenity ubiquity, \textbf{(F)} and neighborhood complexity.}
\label{fig:fig3}
\end{figure}

Figure~\ref{fig:fig3}A, B and C presents amenity diversity, average amenity ubiquity and neighborhood complexity on the map of Budapest, while Figure~\ref{fig:fig3}D, E and F illustrates their correlation with the geographical centrality of neighborhoods in the city. The geographical centrality of locations (both in case of neighborhoods and actual amenities) is determined by the inverse of the average distance to reach the centroid of the location (on a logarithmic scale) from the center of every census tract in Budapest. Since census tracts are relatively homogeneous in terms of population, but heterogeneous in their area, this measure gives us higher values for more densely populated, central locations around the historical city center. To facilitate interpretation, the measure is normalized to a scale of 0-1. As this metric is not based on heuristics or local knowledge, it can be applied to other cities and is motivated by the approach taken by \textcite{moro2021naturecomm}, who showed in detail that the average travel distance to locations is related to the diversity of visitors. In section 7 of the Supplementary information, we provide further details on our measure of geographic centrality, compare our results with the use of several other centrality metrics, and discuss their differences in terms of expectations about their social mixing abilities. Figure~\ref{fig:fig3} suggests that all three variables correlate with central location, but the correlation is stronger for the average amenity ubiquity (-0.538) and neighborhood complexity (-0.464).

\begin{figure}[!ht]
\includegraphics[width=0.915\textwidth]{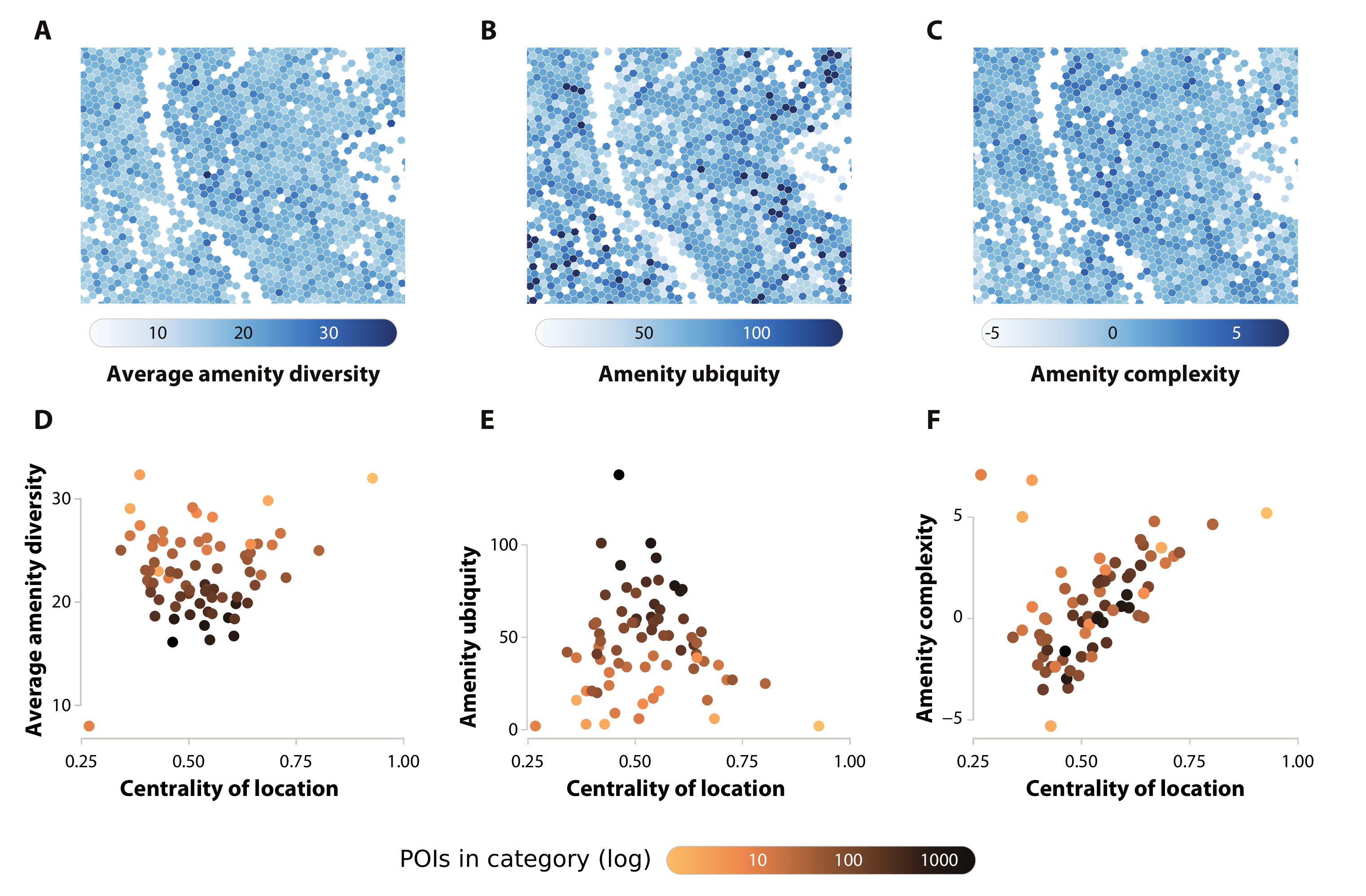}  
\caption{Components of amenity complexity and their association with geographical centrality in Budapest.
\textbf{(A)} Amenities colored by their average amenity diversity, \textbf{(B)} by their amenity ubiquity, \textbf{(C)} and by their amenity complexity in the map of the city center.
\textbf{(D)} Relationship between the geographical centrality of locations and average diversity, \textbf{(E)} ubiquity, and \textbf{(F)} complexity of amenity categories. Each dot is an amenity category and centrality of location is a category average.
}
\label{fig:fig4}
\end{figure}

Figures~\ref{fig:fig4}A, B and C illustrate actual amenities on a zoomed in map of inner Budapest through size 10 H3 hexagons colored by the average diversity, ubiquity and amenity complexity of amenity categories at the location. At dense inner locations of the city, some hexagons contain amenities in multiple amenity categories. The identification of the dominant amenity category is detailed in section 8 of the Supplementary information. Figures~\ref{fig:fig4}D, E and F show how average amenity diversity, amenity ubiquity and amenity complexity are associated with central location. While average diversity and ubiquity of amenities have no clear connection to urban centrality (correlations are 0.180 and -0.056), Figure~\ref{fig:fig4}F suggests that complex amenities tend to be located around the city center (correlation is 0.451, with three remarkable outliers).

\section{Results}

\subsection{Diversity of visitors to complex urban neighborhoods}
\justify

\begin{figure}[!ht]
\includegraphics[width=0.875\textwidth]{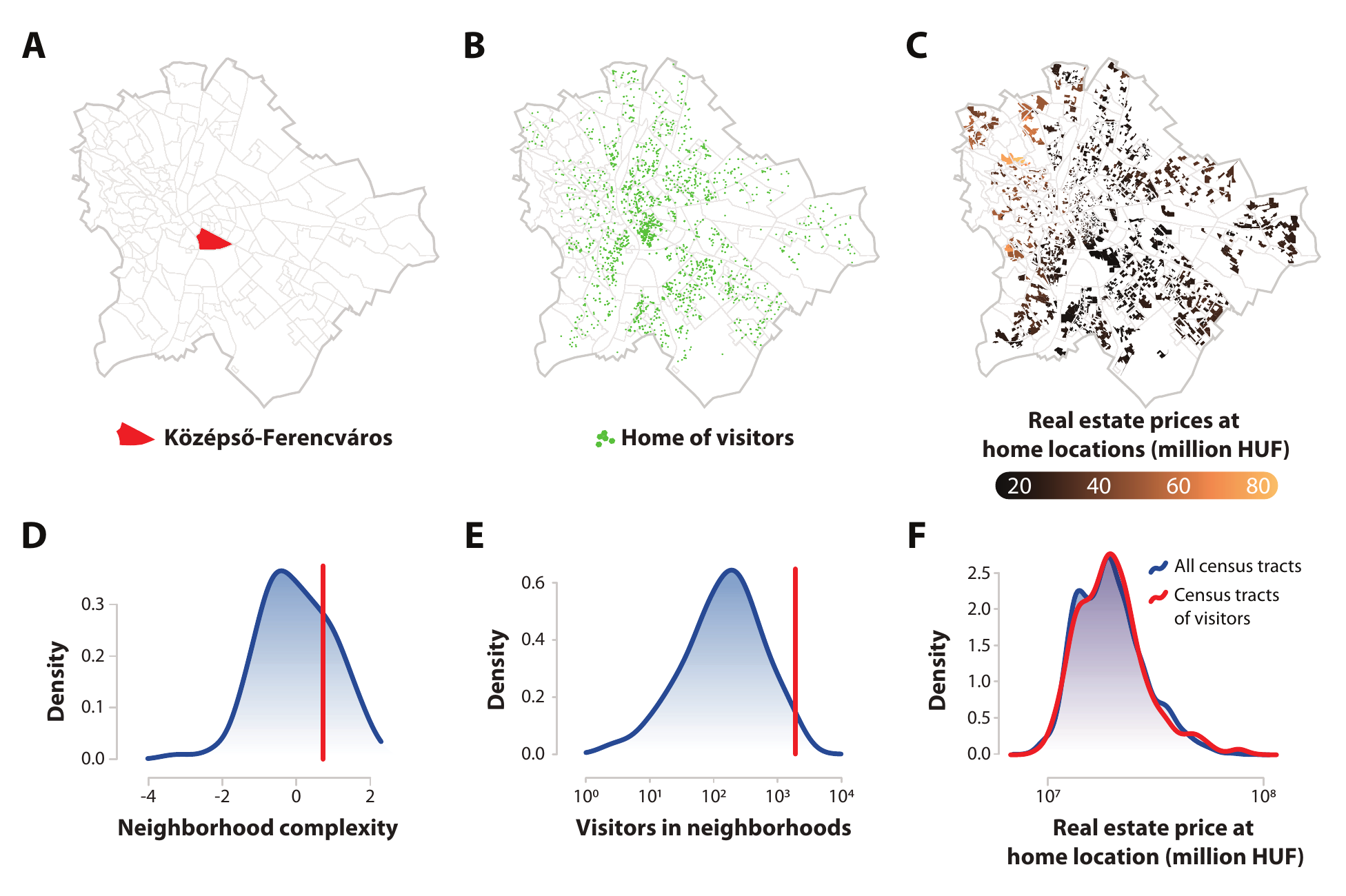}  
\caption{Neighborhood complexity and visitors to an example neighborhood in February 2020.
\textbf{(A)} Selected urban neighborhood of Középső-Ferencváros.
\textbf{(B)} Home location of devices visiting Középső-Ferencváros.
\textbf{(C)} Real estate prices at the home location of visitors.
\textbf{(D)} Distribution of neighborhood complexity values. The red vertical line indicates the complexity of the selected neighborhood of Középső-Ferencváros.
\textbf{(E)} Distribution of observed visitors in neighborhoods. The red vertical line indicates the number of visitors in the selected neighborhood.
\textbf{(F)} Distribution of real estate prices across all census tracts and at the home census tracts of visitors to the selected neighborhood.}
\label{fig:fig5}
\end{figure}

To illustrate the properties of urban locations that attract people of diverse socio-economic status, we combine our neighborhood complexity index with more granular visitation patterns from mobility data. Figure~\ref{fig:fig5} presents our process to join data sources through the example neighborhood of Középső-Ferencváros.

Figure~\ref{fig:fig5}A presents the location of the selected neighborhood, while Figure~\ref{fig:fig5}B visualizes the home location of devices that visited any third places in Középső-Ferencváros during the month of February 2020. We connect the home location of visitors to census tracts as Figure~\ref{fig:fig5}C illustrates. This allows us to infer the socio-economic status of visitors reflected by the real estate prices at the census tract of their home location. Figure~\ref{fig:fig5}D shows that the amenity mix at the selected neighborhood is relatively complex, while Figure~\ref{fig:fig5}E and F show that Középső-Ferencváros is visited by more devices than most neighborhoods in February 2020 and its visitors come from diverse census tracts from all around Budapest.

\renewcommand{\arraystretch}{0.775}
\begin{table}[!ht] \centering 
  \caption{Controlled correlations between the socio-economic diversity of visitors and the amenity complexity of neighborhoods} 
  \label{table:table1} 
\begin{tabular}{@{\extracolsep{5pt}}lccccc} 
\hline \\[-1.8ex] 
\\[-1.8ex] & \multicolumn{5}{c}{Coefficient of variation} \\ 
\\[-1.8ex] & (1) & (2) & (3) & (4) & (5)\\ 
\hline \\[-1.8ex] 
 Neighborhood complexity &  & 0.125$^{***}$ &  &  & 0.125$^{***}$ \\ 
  &  & (0.036) &  &  & (0.038) \\ 
  & & & & & \\ 
 Amenity diversity &  &  & $-$0.148$^{*}$ &  & $-$0.162$^{**}$ \\ 
  &  &  & (0.076) &  & (0.074) \\ 
  & & & & & \\ 
 Avg amenity ubiquity &  &  &  & $-$0.049 & $-$0.017 \\ 
  &  &  &  & (0.049) & (0.050) \\ 
  & & & & & \\ 
 Centrality of location & 0.130$^{***}$ & 0.089$^{***}$ & 0.129$^{***}$ & 0.123$^{***}$ & 0.087$^{***}$ \\ 
  & (0.027) & (0.028) & (0.026) & (0.027) & (0.028) \\ 
  & & & & & \\ 
 Population (log) & $-$0.053$^{**}$ & $-$0.050$^{**}$ & $-$0.049$^{**}$ & $-$0.047$^{**}$ & $-$0.043$^{*}$ \\ 
  & (0.022) & (0.022) & (0.022) & (0.023) & (0.022) \\ 
  & & & & & \\ 
 Nr visitors (log) & 0.047 & 0.039 & 0.054$^{*}$ & 0.045 & 0.046 \\ 
  & (0.029) & (0.029) & (0.029) & (0.029) & (0.029) \\ 
  & & & & & \\ 
 Nr POIs (log) & $-$0.001 & 0.006 & 0.047 & $-$0.013 & 0.054 \\ 
  & (0.023) & (0.023) & (0.033) & (0.026) & (0.034) \\ 
  & & & & & \\ 
 Constant & 0.424$^{***}$ & 0.363$^{***}$ & 0.360$^{***}$ & 0.460$^{***}$ & 0.306$^{***}$ \\ 
  & (0.056) & (0.057) & (0.064) & (0.066) & (0.075) \\ 
  & & & & & \\ 
\hline \\[-1.8ex] 
Observations & 186 & 186 & 186 & 186 & 186 \\ 
R$^{2}$ & 0.257 & 0.303 & 0.273 & 0.261 & 0.321 \\ 
Adjusted R$^{2}$ & 0.241 & 0.284 & 0.253 & 0.241 & 0.295 \\ 
\hline \\[-1.8ex] 
\textit{Note:}  & \multicolumn{5}{r}{$^{*}$p$<$0.1; $^{**}$p$<$0.05; $^{***}$p$<$0.01} \\ 
\end{tabular} 
\end{table}

To capture socio-economic mixing at urban locations, we measure the diversity of visitors in each neighborhood for every month by calculating the coefficient of variation (ratio of standard deviation to the mean) of the real estate prices at the home census tracts of visitors. We focus only on neighborhoods with at least 10 observed visitors in the focal month to get meaningful measures. Table~\ref{table:table1} presents controlled correlations testing the relationship between the diversity of visitors and the amenity structure of neighborhoods using simple OLS regressions on February 2020 data. Model (1) is our baseline model that illustrates the relationship between the diversity of visitors and the centrality of neighborhoods, while controlling for population, number of visitors and number of POIs in neighborhoods. The positive and significant coefficient for urban centrality suggests that central neighborhoods that are on average less distant from the census tracts of Budapest are visited by more diverse people. Model (2) builds on the same model structure, but includes neighborhood complexity as an explanatory variable. It suggests that neighborhood complexity has positive and significant correlation with the diversity of visitors, while taking into account, among other things, the urban centrality of neighborhoods. 
Interestingly, the diversity of amenities shows a mere negative correlation, while the average ubiquity of amenities is not correlated with socio-economic diversity of visitors in our case (see models (3) and (4) in Table~\ref{table:table1}). Model (5) includes all the key explanatory variables and highlight the stable, significant connection between neighborhood complexity and the diversity of visitors.

The average variance inflation factor (VIF) is below 10 in all of the above models, and the VIFs of our main explanatory variables are below 2 in all cases, indicating no serious problems of multicollinearity. In section 7 of the Supplementary information, we illustrate the robustness of our results by using a number of alternative measures of location centrality. Models using centrality metrics based on local knowledge show slightly different results, but for our main models in Table~\ref{table:table1} we use the most general and adoptable centrality measure.
Using the Gini coefficient or the Theil index to capture the diversity of visitors, we get the same results. Related model outputs can be found in section 9 of the Supplementary information. We run the same models presented in Table~\ref{table:table1} on the visitation patterns of non-local users only and observe similar results. In this setting we only consider users living outside the focal neighborhood. Related models are presented in section 10 of the Supplementary information.

Furthermore, the relationship between neighborhood complexity and socio-economic diversity of visitors is estimated for each of the available 24 months using the setting of model (2) in Table~\ref{table:table1}. The related figure in section 11 of the Supplementary information suggests that neighborhood complexity has a positive and significant relationship with the diversity of visitors to neighborhoods in 18 out of the available 24 months. The possible reasons behind the uneven coefficients are discussed in section 5.

Both Figure~\ref{fig:fig3} and Figure~\ref{fig:fig4} suggest that our neighborhood and amenity complexity measures are correlated to urban centrality. Indeed, neighborhood complexity and amenity complexity are derived from the spatial distribution of amenities and then used to explain visits to spatial units, which may raise spatial autocorrelation and endogeneity problems \citep{broekel2019complexity, salinas2021distancecomplexity}. We address potential endogeneity issues by applying two different instrumental variable (IV) approaches. Results of IV regressions presented in section 12 of the Supplementary information further strengthen our argument that neighborhood complexity is connected to the socio-economic diversity of visitors.

\subsection{Diversity of visitors to complex amenities}
\justify

\begin{figure}[!ht]
\includegraphics[width=0.875\textwidth]{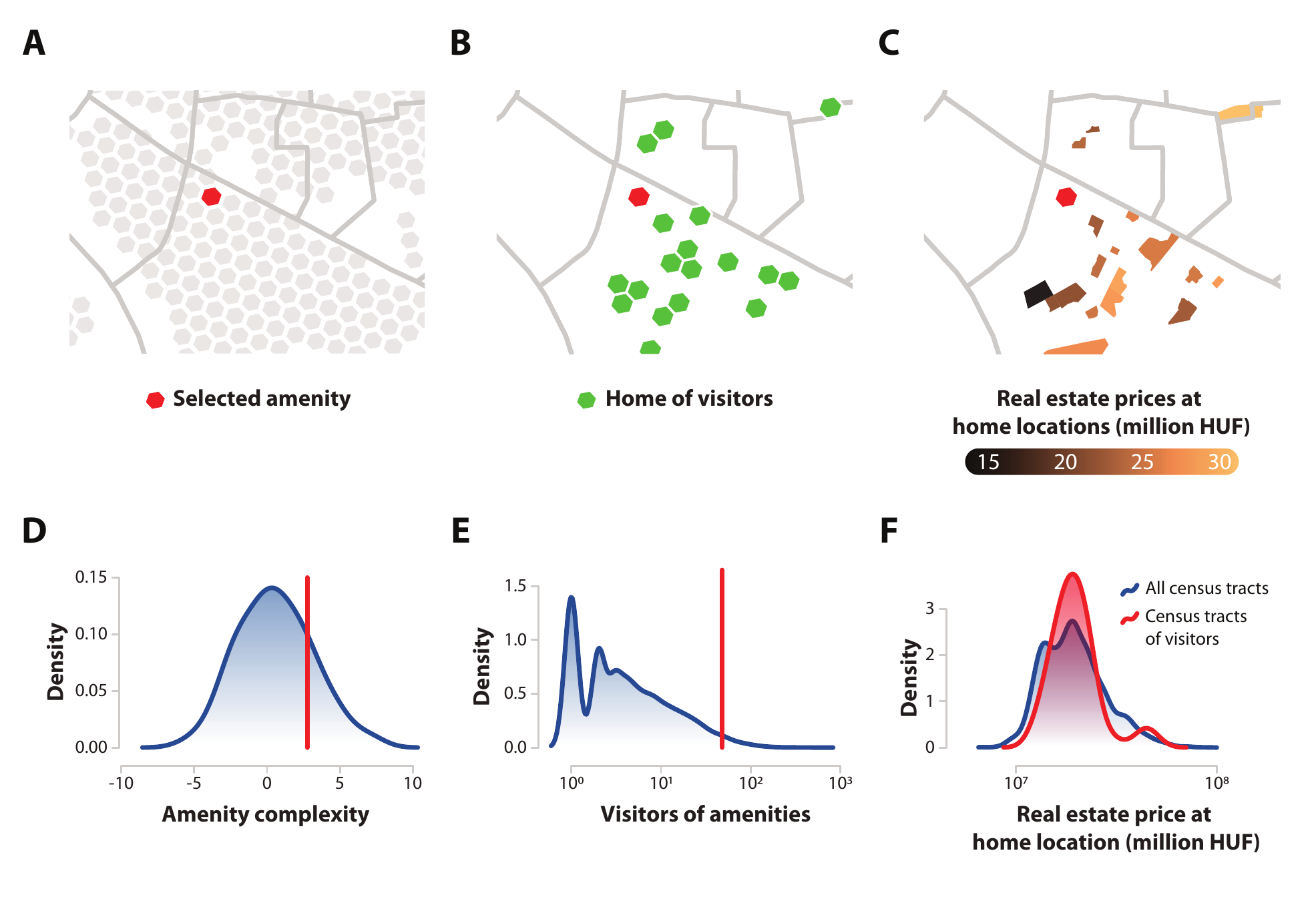}  
\caption{
Connecting amenity complexity to visitor diversity. 
\textbf{(A)} Selected bar on a map. Light red color hexagons indicate other nearby amenities.
\textbf{(B)} Neighboring home location of visitors.
\textbf{(C)} Real estate prices in the census tract of the visitor home locations.
\textbf{(D)} Distribution of amenity complexity values. The red vertical line indicates the amenity complexity of bars, the selected amenity category.
\textbf{(E)} Distribution of visitors to observed amenities in February 2020, Budapest. The red vertical line indicates the number of visitors at the example bar.
\textbf{(F)} Distribution of real estate prices in all census tracts and in census tracts where the visitors of the example bar live.
}
\label{fig:fig6}
\end{figure}

To go beyond the level of neighborhoods, we combine amenity complexity measured at the amenity category level with visitations to actual amenities derived from our fine-grained mobility data. Figure~\ref{fig:fig6} presents our process to join data sources at the level of amenities through an example bar in the neighborhood of Középső-Ferencváros, Budapest. The selected amenity is surrounded by other amenities (Figure~\ref{fig:fig6}A) and by detecting the home location of visitor devices (in Figure~\ref{fig:fig6}B we use February 2020 data and the surrounding area in size 10 H3 hexagons), we can observe the socio-economic status of visitors proxied by real estate prices (Figure~\ref{fig:fig6}C). Our Bar example is a relatively complex amenity category (Figure~\ref{fig:fig6}D) and is very frequently visited in comparison to other observed amenities in February 2020 (Figure~\ref{fig:fig6}E). In addition, visitors from census tracts with medium or higher real estate prices are over-represented in February 2020, as shown in Figure~\ref{fig:fig6}F.

We measure the socio-economic diversity of visitors to each amenity for every month by calculating the coefficient of variation (ratio of standard deviation to the mean) of the real estate prices at the home census tracts of visitors. To do so, we focus only on amenities with at least 10 observed visitors in the focal month that helps us avoid meaningless values of the indicator. Table~\ref{table:table2} presents simple OLS models to illustrate the relationship between the socio-economic diversity of visitors and components of amenity complexity at the level of amenities in February 2020. All our models at the level of amenities present clustered standard errors at the level of amenity categories.

We find in Model (1) that geographical centrality is a significant predictor of socio-economic mixing at amenities, even after controlling for the total number of POIs in the respective amenity category across Budapest and the observed number of visitors to the amenity in the focal month. Model (2) presents that amenity complexity has an additional positive and significant relationship with the socio-economic mixing. The negative and significant coefficient of amenity ubiquity in Model (3) suggests that amenities that many urban locations are specialized in are visited by less diverse, while amenities that only few neighborhoods specialized in are visited by more diverse groups of people.
This result is consistent with the findings of \textcite{moro2021naturecomm}. The positive and slightly significant coefficient on the average diversity of amenities in Model (4) indicates that amenity categories that mostly appear in diverse neighborhoods attract visitors with different socio-economic status. Model (5) includes all the key explanatory variables and highlight the stable, significant connection between amenity complexity and the diversity of visitors. While the effect of amenity diversity remains stable, the significance of average amenity diversity disappears in the final model.

\renewcommand{\arraystretch}{0.775}
\begin{table}[!hb] \centering 
  \caption{Controlled correlations between the socio-economic diversity of visitors and amenity complexity} 
  \label{table:table2} 
\begin{tabular}{@{\extracolsep{5pt}}lccccc} 
\hline \\[-1.8ex] 
\\[-1.8ex] & \multicolumn{5}{c}{Coefficient of variation} \\ 
\\[-1.8ex] & (1) & (2) & (3) & (4) & (5)\\ 
\hline \\[-1.8ex] 
 Amenity complexity &  & 0.069$^{***}$ &  &  & 0.040$^{***}$ \\ 
  &  & (0.016) &  &  & (0.015) \\ 
  & & & & & \\ 
 Amenity ubiquity &  &  & $-$0.071$^{***}$ &  & $-$0.051$^{**}$ \\ 
  &  &  & (0.015) &  & (0.018) \\ 
  & & & & & \\ 
 Avg amenity diversity &  &  &  & 0.047$^{**}$ & $-$0.007 \\ 
  &  &  &  & (0.024) & (0.018) \\ 
  & & & & & \\ 
 Centrality of location & 0.126$^{***}$ & 0.116$^{***}$ & 0.118$^{***}$ & 0.125$^{***}$ & 0.115$^{***}$ \\ 
  & (0.011) & (0.011) & (0.010) & (0.011) & (0.010) \\ 
  & & & & & \\ 
 Nr POIs in category (log) & $-$0.013 & $-$0.006 & 0.015$^{*}$ & $-$0.001 & 0.009 \\ 
  & (0.009) & (0.007) & (0.009) & (0.009) & (0.008) \\ 
  & & & & & \\ 
 Nr visitors (log) & 0.073$^{***}$ & 0.069$^{***}$ & 0.068$^{***}$ & 0.071$^{***}$ & 0.067$^{***}$ \\ 
  & (0.007) & (0.008) & (0.008) & (0.007) & (0.008) \\ 
  & & & & & \\ 
 Constant & 0.220$^{***}$ & 0.172$^{***}$ & 0.180$^{***}$ & 0.173$^{***}$ & 0.170$^{***}$ \\ 
  & (0.033) & (0.030) & (0.027) & (0.036) & (0.028) \\ 
  & & & & & \\ 
\hline \\[-1.8ex] 
Observations & 2,742 & 2,742 & 2,742 & 2,742 & 2,742 \\ 
R$^{2}$ & 0.102 & 0.108 & 0.108 & 0.104 & 0.110 \\ 
Adjusted R$^{2}$ & 0.101 & 0.106 & 0.107 & 0.102 & 0.108 \\ 
\hline \\[-1.8ex] 
\textit{Note:}  & \multicolumn{5}{r}{$^{*}$p$<$0.1; $^{**}$p$<$0.05; $^{***}$p$<$0.01} \\ 
 & \multicolumn{5}{r}{Standard errors are clustered at the amenity category level.} \\ 
\end{tabular} 
\end{table}

The average VIF value is below 3 in all of the above models, and the VIFs of our main explanatory variables are close 1 in all cases, indicating no serious problems of multicollinearity. Results with alternative measures for location centrality can be found in section 7 of the Supplementary information. Amenity complexity is significantly related to socio-economic mixing using almost any centrality indicators. Changing the dependent variable to the Gini coefficient or the Theil index to capture the diversity of visitors to amenities, we obtain the same results. Related model outputs can be found in section 9 of the Supplementary information. The relationship between amenity complexity and socio-economic diversity of visitors is estimated for each of the available 24 months using the setting of model (2) in Table~\ref{table:table2}. The related figure in section 11 in the Supplementary information presents that amenity complexity has a positive and significant relationship with the diversity of visitors to amenities in 16 out of the available 24 months.

\section{Discussion}
\justify

In this work we bring the ideas behind economic complexity metrics to the urban problems of experienced segregation and social mixing. We measure amenity complexity by utilizing the spatial distribution of point of interests (POIs) inside a city. We combine the information on the complexity of neighborhoods and of amenity categories with fine-grained mobility data to illustrate the relationship between the complexity of amenities available in a location and the socio-economic diversity of its visitors.
Focusing on the urban neighborhoods of Budapest, Hungary, we find that neighborhoods that concentrate a more complex amenity mix attract a bigger diversity of socio-economic groups. Applying the same logic to actual amenities inside Budapest, we also show that POIs of more complex amenity categories are visited by larger diversities of strata. However, the diversity and ubiquity of amenities, the two components of amenity complexity, show a less clear relationship with the socio-economic diversity of visitors. Diversity of amenities shows a surprising negative correlation with visitor diversity, but only at the neighborhood level, while ubiquity of amenities is related to visitor diversity only at the amenity category level.

The geographical centrality of urban locations is a strong predictor of socio-economic mixing. Our results illustrate that both neighborhood complexity and amenity complexity correlate with the geographical centrality of locations. Contrary to previous works \citep{zhong2015centrality}, we find that diversity of amenities is less correlated to urban centrality, while amenity ubiquity is only associated with centrality of locations at the neighborhood level. The relationship between amenity complexity and centrality of locations has inspired a number of robustness checks, including the use of instrumental variable regressions (details can be found in section 7 and section 12 of the Supplementary information), which further confirm our key finding that amenity complexity is associated with socio-economic diversity of visitors.

% contribution
The general contribution of our paper is that we combine economic complexity concepts with urban mobility research. Constructing the measures of neighborhood and amenity complexity allows us to systematically test the contribution of certain amenity categories and also the amenity portfolio at certain locations to socio-economic mixing in cities. Moreover, we contribute to the line of research on segregation patterns inside cities by illustrating in a direct fashion based on fine-grained mobility data that centrality of urban locations largely influence socio-economic mixing.

% limitations
Our empirical work has several limitations, but offers promising future research directions. This study only focuses on the city of Budapest. Budapest is the only large city in Hungary and it clearly has a monocentric structure. Therefore, our findings are limited to this specific context and similar empirical works in cities with different size, geography and urban structure are necessary to assess the generality of our conclusions.

To construct the amenity complexity measures, we rely on the specific neighborhood structure of Budapest, however, alternative spatial scales in different urban settings are necessary to be tested in the future. We believe that the level of neighborhoods is the appropriate spatial scale to construct amenity complexity metrics for two reasons. First, the size of the applied spatial units can influence the nestedness of the location-amenity matrix used to construct complexity metrics. Co-occurrence of POIs in different amenity categories are less likely in case we consider smaller geographical areas. Neighborhoods are proved to be large enough to produce intuitive results. Second, neighborhoods are very important spatial units of urban life. They are argued to be the environment that can influence social capital accumulation and social mobility \citep{chetty2022nature, chetty2016neighborhood}. Moreover, they have clear administrative boarders and people can identify with them, which makes the interpretation of amenity complexity results more appealing. 

Inside Budapest, we observe that central location is correlated to both neighborhood complexity and amenity complexity and these factors are all connected to the socio-economic diversity of visitors. To understand this relationship clearly, we test several different measures of central location, with mixed results. For our main empirical exercise we choose the centrality metric based on the average distance to reach neighborhoods and amenities from any census tracts inspired by \textcite{moro2021naturecomm}, as this measure does not require local knowledge and is easy to adopt for other cities. However, further research is needed to better understand these relationships, as a recent work based on similar measures adopted to Paris, France, shows that amenity complexity is not exclusively linked to a single city center \citep{robertson2023unpacking}.

Mobility data for our empirical analysis are produced on a monthly basis. Our findings are valid for 16 of the 24 months available and COVID-19 did not clearly affect the relationship between amenity complexity and the diversity of visitors to urban locations. However, our data does not contain enough observations to provide significant results for all the 24 months under study. Confirming the results with better mobility data is an important future research direction.

The counting of POIs within neighborhoods does not allow differentiation between the capacity or quality of amenities. We have tested unsuccessfully the number of visitors to amenity categories within neighborhoods as an input rather than the sheer number of POIs. In the future, similar but more sophisticated data would be needed to measure amenity complexity more accurately.

In our empirical exercise, we adopted the most commonly used economic complexity indicator to amenities and neighborhoods. However, several modifications have been suggested to improve economic complexity measurement \citep{tacchella2012fitness, mealy2019interpreting} and the adoption of these methods to the neighborhood scale in urban environments is an apparent future research direction.

\section*{Acknowledgements}
    Sándor Juhász worked on the paper as a Marie Skłodowska-Curie Postdoctoral Fellow at the Complexity Science Hub Vienna (grant number 101062606). The work of Balázs Lengyel was financially supported by Hungarian National Scientific Fund (OTKA K 138970). The authors acknowledge the help of Orsolya Vásárhelyi and Luis Guillermo Natera Orozco with the original POI data collection. We wish to thank Tom Broekel, Frank Neffke, Gergő Tóth and László Czaller for their comments and suggestions. We acknowledge the significant assistance of our illustrator Szabolcs Tóth-Zs. in finalizing our primary figures. The authors thank the Social Science Computing Unit Budapest, and the Data Bank of the Centre for Economic- and Regional Studies for the contribution in data management.

\newpage
\printbibliography
\newpage

\clearpage

\section*{Supplementary information}

\section*{S1 Mobility data preparation}

Our GPS based mobility data is provided by a data aggregator company that collects and combines anonymous location data from users' smartphone applications. The sample we use for our analysis is an unbalanced panel of GPS pings from 5.2 million devices in Hungary between 2019 June and 2021 May. Our raw data consists of a device identifier, a time stamp, a latitude, and a longitude coordinate, where GPS pings occur. Pings are logged in case an application on a device requests location information. Sometimes this is the result of an active behavior such as using a navigation application or requesting local weather information. In other cases, pings could be the result of an application requesting information while running in the background. As a consequence, pings occur at irregular intervals. To identify visitation patterns such as home, work or third place visits to urban locations, we transform and filter our raw GPS trajectory data in several steps. We detail these steps in the following.

As the data contains some pings attributed to the same device that indicate unreasonable behavior, we start by iteratively removing all ping pairs that signal a movement over 300 kilometers per hour speed. Additionally, we discard all devices that have fewer than 20 pings remaining after this initial speed-based filter. These two steps reduce the total ping count from 3.18 billion to 3.13 billion and the total unique device count from 5.2 million to 1.88 million.

To focus on locations where devices stopped for some time, we run the Infostop stop detection algorithm (\cite{aslak2020infostop}). In short, it classifies GPS pings to trips or stays and clusters stationary points into stops in an effective way. We apply the algorithm with the following parameter set. $r1$, the maximum roaming distance allowed for two pings within the same stay is set to 370 meters. $r2$, the typical distance between two stays in the same destination is set to 140 meters. $t_{min}$, the minimum duration of a stay is set to 270 seconds, while $t_{max}$, the maximum time difference between two consecutive pings to be considered within the same stay is set to 7200 seconds. The minimum number of GPS pings required for stationary points is set to 2.

The parameters are calibrated using the Google location history data of 7 consenting individuals from Budapest, Hungary. We ran our stop detection algorithm on the sample trajectories with a wide range of parameters. We compared the results of the stop detection process to the personal, anecdotal experiences and to the semantic stop detection extracted from Google accounts. This experiment confirmed that the stop detection algorithm and the parameters produce a reasonable set of trips, stays, and destinations. Moreover, the subsequent home and work detection process built on it produced accurate results for our small sample.

Using the output of the stop detection process, we further filter the data to devices that have at least 2 unique destinations with over 4 different stays in each. This reduces the total stop count from 100 million to about 80 million, and the unique device count from 1.75 million to about 240.000. However, even in this final form, over 2.1 billion pings are used from the original 3.18 billion. Our empirical exercise in the end only uses information for each month from devices with identified home, work and at least a single visited third place inside Budapest.

\section*{S2 Socio-economic status from census tract level real estate prices}

We infer on the socio-economic status of individuals living in Budapest by connecting their identified home location to residential real estate prices at the census tract level. Approximating socio-economic status through real estate prices has several benefits in comparison to the prevalent solution of using household income statistics of urban locations. First, real estate prices are by definition connected to places, while it is harder to connect income to locations. Second, real estate statistics in the census are comprehensive, while income information only reflects on the status of active employees.

The Hungarian Central Statistical Office collects data on all residential real estate sales contracts and derives information on transaction prices for the entire country. As not every real estate is on the market and observed contracts sometimes suffer from missing information on the parameters of given properties, direct measurement on lower geographical level is difficult. By utilizing the fact that real estate prices tend to follow a strong multi-level hierarchy as location (and especially neighborhoods in Budapest) explains much of the price differences, we train a multi-level random slope regression model on the observed transaction prices \citep{chi2021shedding, snijders2011multilevel}. To do so, we use real estate transaction contracts between 2013 and 2019. We create a pooled setting by correcting prices through the city level house price index published yearly by the National Bank of Hungary. Our model can be written as:

\begin{equation}
\begin{split}
\label{eq:si_eq1}
    h_{i,j} = \beta_{0,j} + \beta_{1,j}s_{i,j} + \varepsilon_{i,j} \\
    \beta_{0,j} = \beta_{0} + n_{0,j} \\
    \beta_{1,j} = \beta_{1} + n_{1,j}
\end{split}
\end{equation}

\noindent
Here $h_{i,j}$ is the logarithm of the individual price of real estate $i$ in neighborhood $j$. $\beta_{0,j}$ represents how much the estimated mean house price differs by neighborhoods. Estimated mean neighborhood prices are decomposed to $\beta_{0}$, the city level mean, and $n_{0,j}$, the neighborhood deviation from this value. To be able to capture differences within neighborhoods, we apply the individual level parameter $s_{i,j}$ that refers to the size (floor area) of real estate $i$ in neighborhood $j$. Since the effect of floor area can vary between neighborhoods, we train a random slope model. $\beta_{1,j}$ represents the effect of floor area in neighborhood $j$. $\beta_{1,j}$ is decomposed to a city level slope $\beta_{1}$, and the deviation of neighborhood slopes around this value $n_{1,j}$.

\begin{figure}[ht!]
\centering
\includegraphics[width=0.95\textwidth]{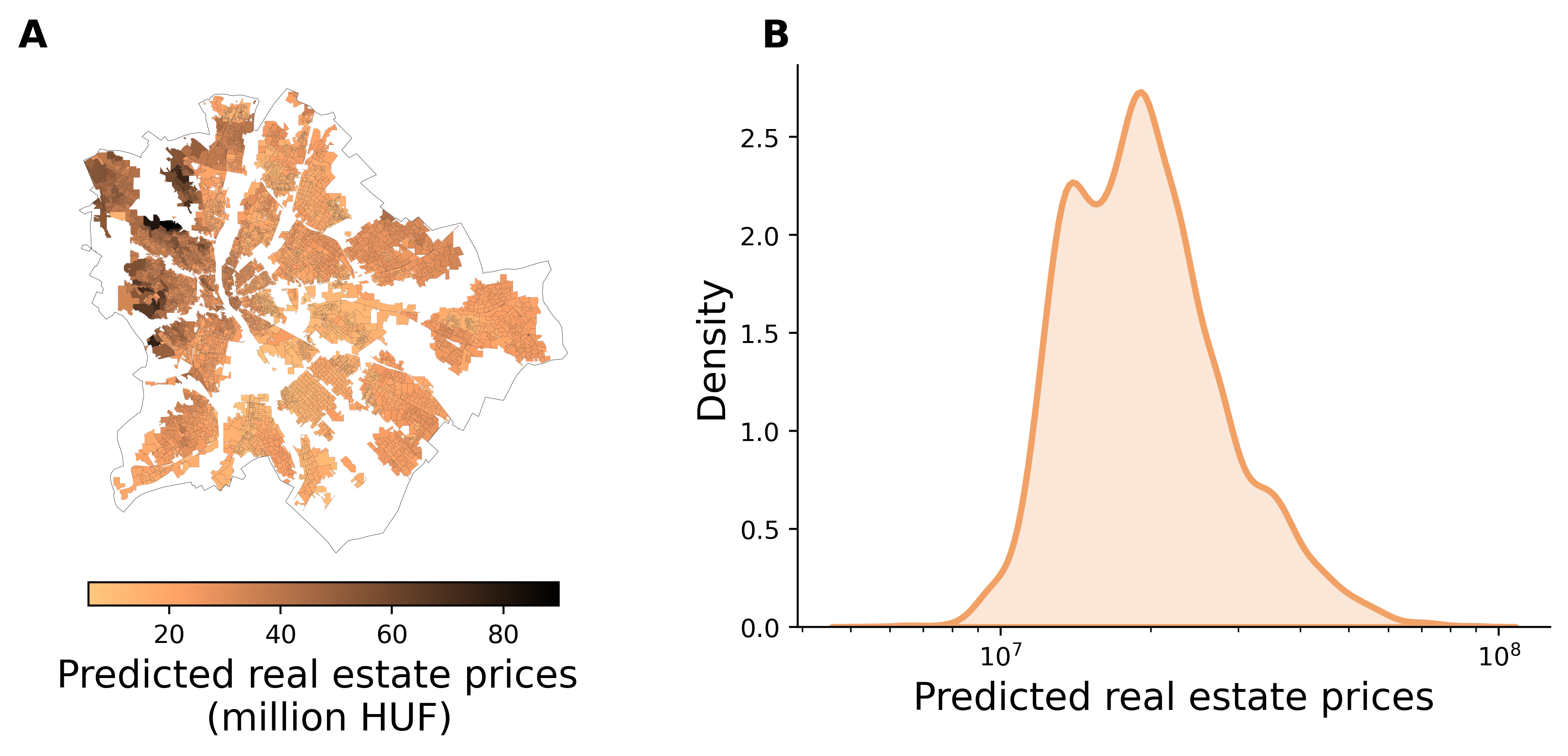}
\caption{Predicted real estate prices at the census tract level around Budapest (A) and the distribution of real estate prices aggregated to the census tract level in Budapest (B)}
\label{fig:si2_fig1}
\end{figure}

Utilizing this model, we predict prices for every real estate captured by the last Hungarian census in 2010. By taking the mean of the predicted real estate prices at the census tract level, we get a highly granular socio-economic status map for the entire city of Budapest. Figure \ref{fig:si2_fig1}A illustrates the predicted prices aggregated to the census tract level on the map of Budapest, while Figure \ref{fig:si2_fig1}B shows the distribution of predicted real estate prices.

\clearpage
\section*{S3 Urban neighborhoods of Budapest, Hungary}

\begin{figure}[htb]
\centering
\includegraphics[width=0.95\textwidth]{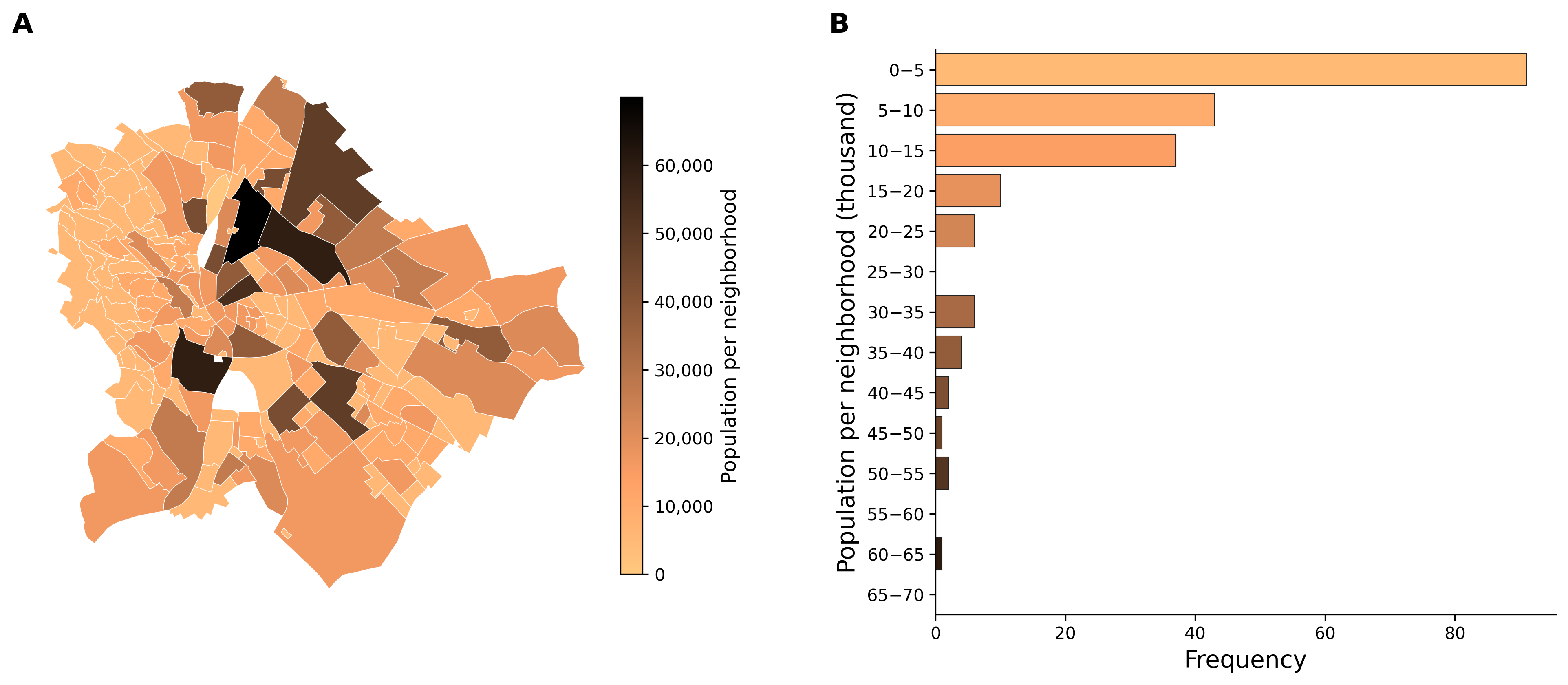}
\caption{Population of urban neighborhoods based on the census of 2010 on the map of Budapest (A) and as a distribution plot (B).}
\label{fig:si3_fig1}
\end{figure}

\begin{figure}[htb]
\centering
\includegraphics[width=0.95\textwidth]{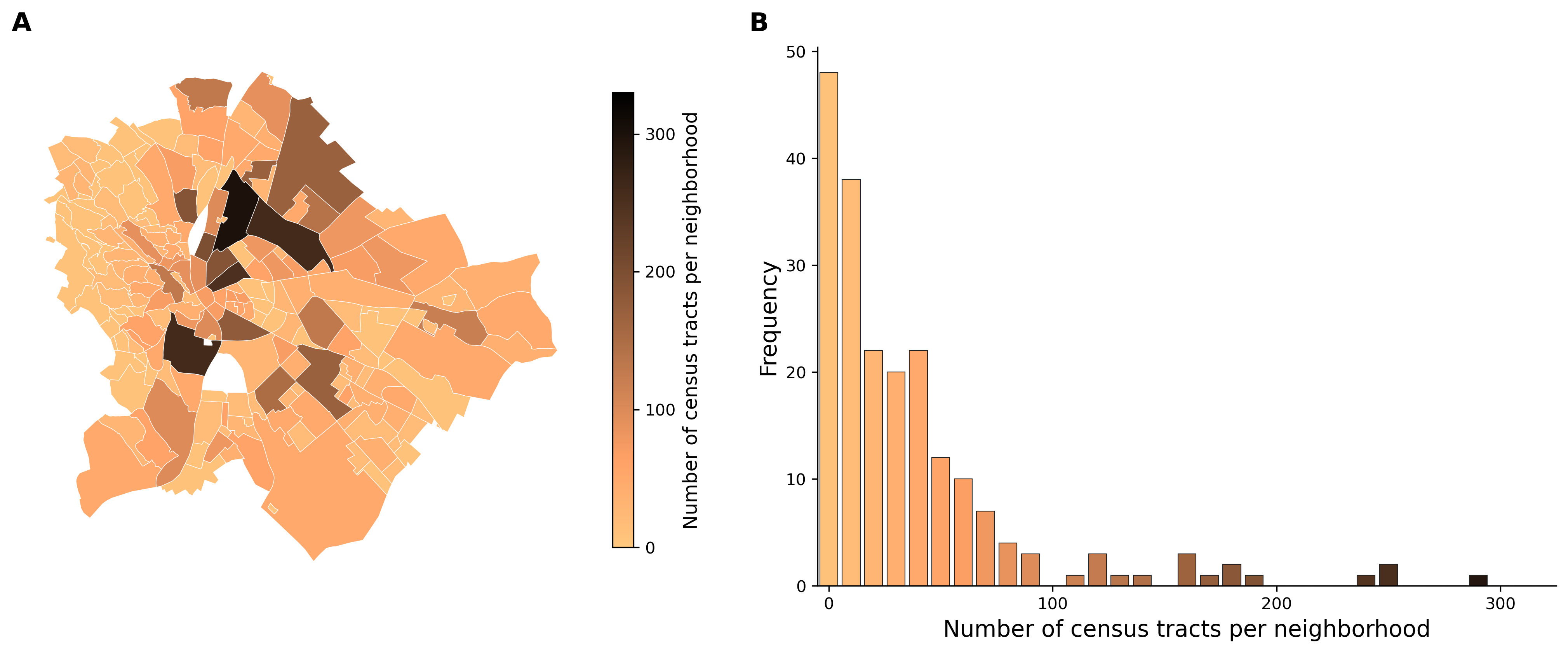}
\caption{Number of census tracts per urban neighborhood based on the census of 2010 on the map of Budapest (A) and as a distribution plot(B).}
\label{fig:si3_fig2}
\end{figure}

We use urban neighborhoods as geographic units to construct our amenity complexity measures. Budapest consists of 207 urban neighborhoods and their place in the spatial scale hierarchy is between districts and census tracts in terms of area and population. The unequal size distribution of urban neighborhoods are illustrated in Figure~\ref{fig:si3_fig1} and Figure~\ref{fig:si3_fig2}. The correlation between population of urban neighborhoods and the number of census tracts per urban neighborhoods is strong as Pearson's R is \num{0.979}. Further details about neighborhoods can be found at the website of \citep{kshregionalatlas}.

\clearpage
\section*{S4 Alternative specifications to construct amenity complexity measures}

\begin{figure}[htb]
\centering
\includegraphics[width=0.9\textwidth]{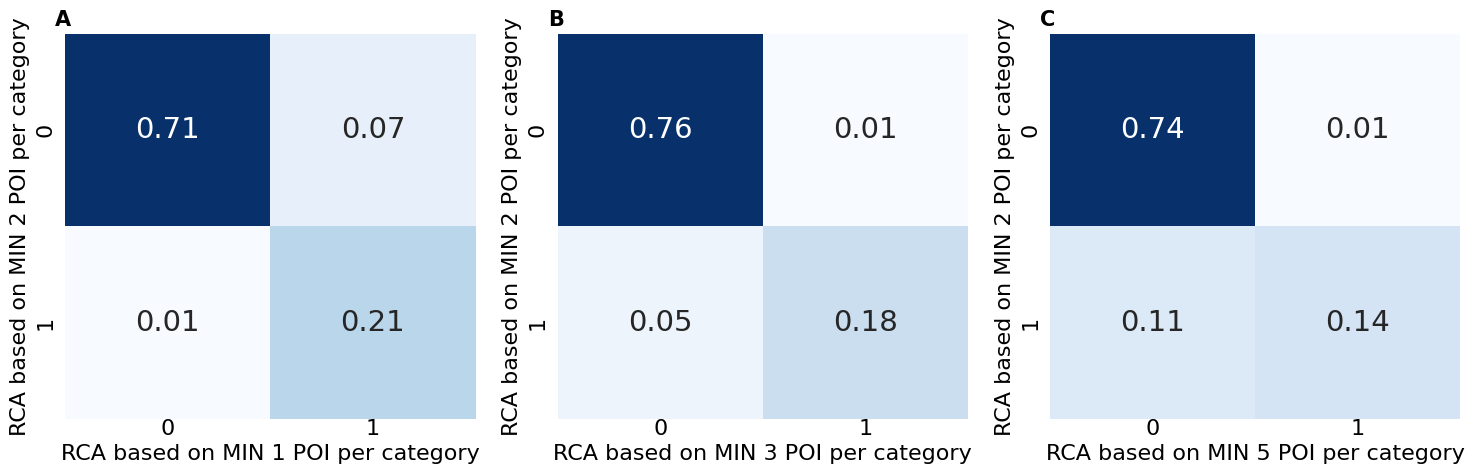}
\caption{Stability of RCA values in the binary specialization matrix ($M$) by different specifications. (A) Changes in RCA 0/1 values in case we consider every amenity category in each neighborhood with at least 1 POI. (B) Changes in RCA 0/1 values in case we increase the minimum number of POIs per amenity category in neighborhoods from 2 to 3. (C) Changes in RCA 0/1 values in case we increase the minimum number of POIs per amenity category in neighborhoods from 2 to 5.}
\label{fig:si4_fig1}
\end{figure}

To construct our amenity complexity measures, we create a neighborhood-amenity category matrix, where we consider every neighborhood in Budapest with at least 2 amenity categories with minimum 2 POIs. The minimum 2 POIs threshold is implemented to reduce noise in our data. We use this matrix to measure the revealed comparative advantage (RCA) of neighborhoods in amenity categories and transform it to a binary specialization matrix ($M$) for the calculation of complexity values at the level of neighborhoods and amenity categories, as described in the main text. Figure~\ref{fig:si4_fig1} presents how the composition of $M$ would change in case we use alternative thresholds for minimum number of POIs per amenity categories in neighborhoods.

\begin{figure}[!ht]
\centering
\includegraphics[width=0.75\textwidth]{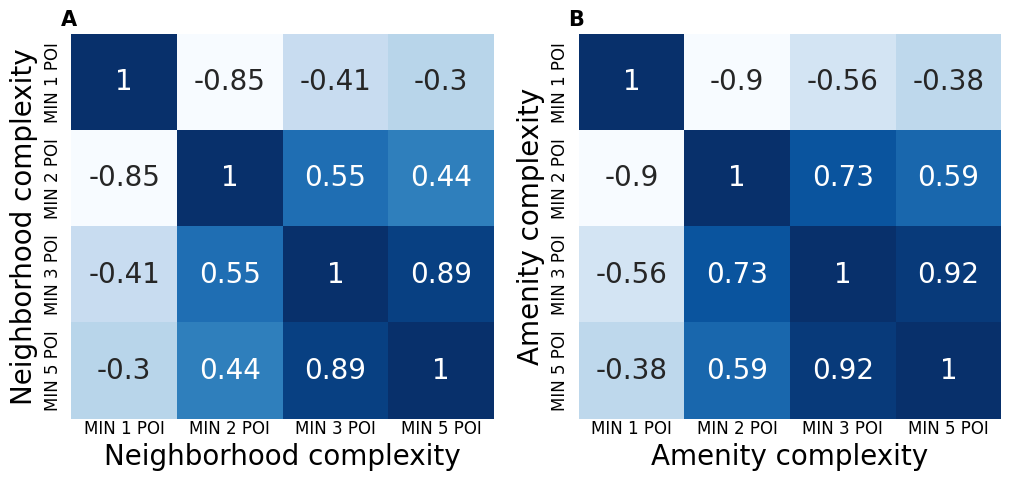}
\caption{Correlation tables for (A) neighborhood complexity and (B) amenity complexity values created by applying different minimum thresholds on the number of POIs in the neighborhood-amenity category matrix.}
\label{fig:si4_fig2}
\end{figure}

\renewcommand{\arraystretch}{0.95}
\begin{table}[!ht] \centering 
  \caption{Controlled correlations between the diversity of visitors and the complexity of neighborhoods (in February 2020) in cases we apply different minimum thresholds for the number of POIs considered in the neighborhood-amenity category matrix.} 
  \label{table:si4_table1} 
\begin{tabular}{@{\extracolsep{5pt}}lccc} 
\hline \\[-1.8ex] 
\\[-1.8ex] & \multicolumn{3}{c}{Coefficient of variation} \\ 
\\[-1.8ex] & (1) & (2) & (3)\\ 
\hline \\[-1.8ex] 
 Neighborhood complexity & 0.125$^{***}$ &  &  \\ 
  & (0.036) &  &  \\ 
  & & & \\ 
 Neighborhood complexity min 3 POIs &  & 0.133$^{***}$ &  \\ 
  &  & (0.045) &  \\ 
  & & & \\ 
 Neighborhood complexity min 5 POIs &  &  & 0.098$^{**}$ \\ 
  &  &  & (0.039) \\ 
  & & & \\ 
 Centrality of location & 0.089$^{***}$ & 0.102$^{***}$ & 0.117$^{***}$ \\ 
  & (0.028) & (0.028) & (0.027) \\ 
  & & & \\ 
 Population (log) & $-$0.050$^{**}$ & $-$0.031 & $-$0.021 \\ 
  & (0.022) & (0.023) & (0.023) \\ 
  & & & \\ 
 Nr visitors (log) & 0.039 & 0.037 & 0.040 \\ 
  & (0.029) & (0.029) & (0.029) \\ 
  & & & \\ 
 Nr POIs (log) & 0.006 & $-$0.039 & $-$0.052$^{*}$ \\ 
  & (0.023) & (0.025) & (0.028) \\ 
  & & & \\ 
 Constant & 0.363$^{***}$ & 0.394$^{***}$ & 0.396$^{***}$ \\ 
  & (0.057) & (0.055) & (0.055) \\ 
  & & & \\ 
\hline \\[-1.8ex] 
Observations & 186 & 184 & 177 \\ 
R$^{2}$ & 0.303 & 0.275 & 0.255 \\ 
Adjusted R$^{2}$ & 0.284 & 0.255 & 0.233 \\ 
\hline 
\textit{Note:}  & \multicolumn{3}{r}{$^{*}$p$<$0.1; $^{**}$p$<$0.05; $^{***}$p$<$0.01} \\ 
\end{tabular} 
\end{table}

\renewcommand{\arraystretch}{0.95}
\begin{table}[!ht] \centering 
  \caption{Controlled correlations between the diversity of visitors and the complexity of amenities (in February 2020) in cases we apply different minimum thresholds for the number of POIs considered in the neighborhood-amenity category matrix.} 
  \label{table:si4_table2}
\begin{tabular}{@{\extracolsep{5pt}}lccc} 
\hline \\[-1.8ex] 
\\[-1.8ex] & \multicolumn{3}{c}{Coefficient of variation} \\ 
\\[-1.8ex] & (1) & (2) & (3)\\ 
\hline \\[-1.8ex] 
 Amenity complexity & 0.069$^{***}$ &  &  \\ 
  & (0.016) &  &  \\ 
  & & & \\ 
 Amenity complexity min 3 POIs &  & 0.077$^{***}$ &  \\ 
  &  & (0.020) &  \\ 
  & & & \\ 
 Amenity complexity min 5 POIs &  &  & 0.094$^{***}$ \\ 
  &  &  & (0.030) \\ 
  & & & \\ 
 Centrality of location & 0.116$^{***}$ & 0.118$^{***}$ & 0.116$^{***}$ \\ 
  & (0.011) & (0.011) & (0.011) \\ 
  & & & \\ 
 Nr POIs in category (log) & $-$0.006 & 0.012 & 0.022$^{*}$ \\ 
  & (0.007) & (0.009) & (0.012) \\ 
  & & & \\ 
 Nr visitors (log) & 0.069$^{***}$ & 0.067$^{***}$ & 0.067$^{***}$ \\ 
  & (0.008) & (0.008) & (0.008) \\ 
  & & & \\ 
 Constant & 0.172$^{***}$ & 0.121$^{***}$ & 0.094$^{**}$ \\ 
  & (0.030) & (0.036) & (0.045) \\ 
  & & & \\ 
\hline \\[-1.8ex] 
Observations & 2,742 & 2,740 & 2,739 \\ 
R$^{2}$ & 0.108 & 0.107 & 0.108 \\ 
Adjusted R$^{2}$ & 0.106 & 0.106 & 0.106 \\ 
\hline \\[-1.8ex] 
\textit{Note:}  & \multicolumn{3}{r}{$^{*}$p$<$0.1; $^{**}$p$<$0.05; $^{***}$p$<$0.01} \\ 
\end{tabular} 
\end{table}

Despite the different thresholds, around 90\% of the neighborhood-amenity category matrix is classified by the same 0/1 RCA value. This suggests that matrix $M$ is relative stable for smaller changes in the minimum number of POIs considered. In case we do not introduce any threshold to construct the matrix $M$, we classify slightly more neighborhood-amenity category with RCA=1.

Figure~\ref{fig:si4_fig2} presents the correlation tables for neighborhood and amenity complexity values based on different minimum thresholds on the number of POIs. The tables suggest that there is a clear difference in complexity values in case we introduce a minimum threshold and in case we do not. Without any noise filtering the resulted complexity values have a negative correlation to complexity values constructed from neighborhood-amenity pairs with at least 2 POIs. We opt to choose the minimum 2 POIs threshold as it results in intuitive complexity values for both neighborhoods and amenity categories, but still not too restrictive. Table~\ref{table:si4_table1} and Table~\ref{table:si4_table2} presents that our main results are not influenced by increasing the minimum 2 POIs threshold to minimum 3 or 5 POIs.

\begin{figure}[!ht]
\centering
\includegraphics[width=0.85\textwidth]{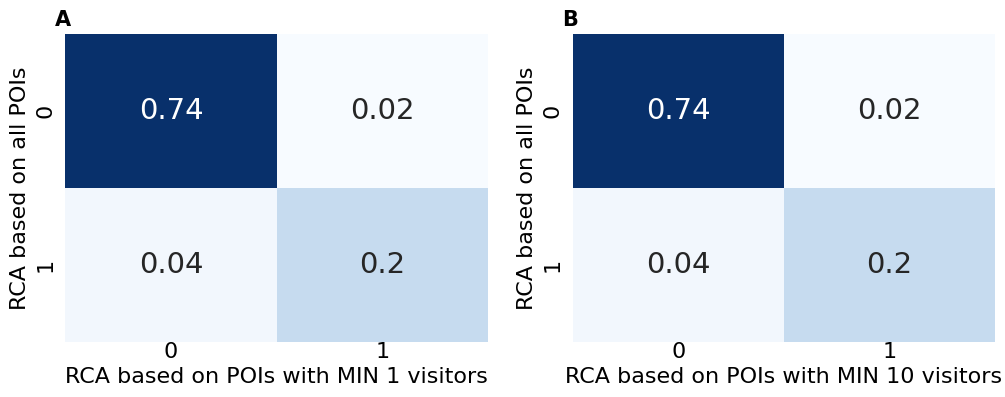}
\caption{Stability of RCA values in the binary specialization matrix ($M$) in case we only consider POIs (A) visited by at least 1 device, (B) visited by at least 10 devices during the observed 24 months.}
\label{fig:si4_fig3}
\end{figure}

\begin{figure}[!ht]
\centering
\includegraphics[width=0.75\textwidth]{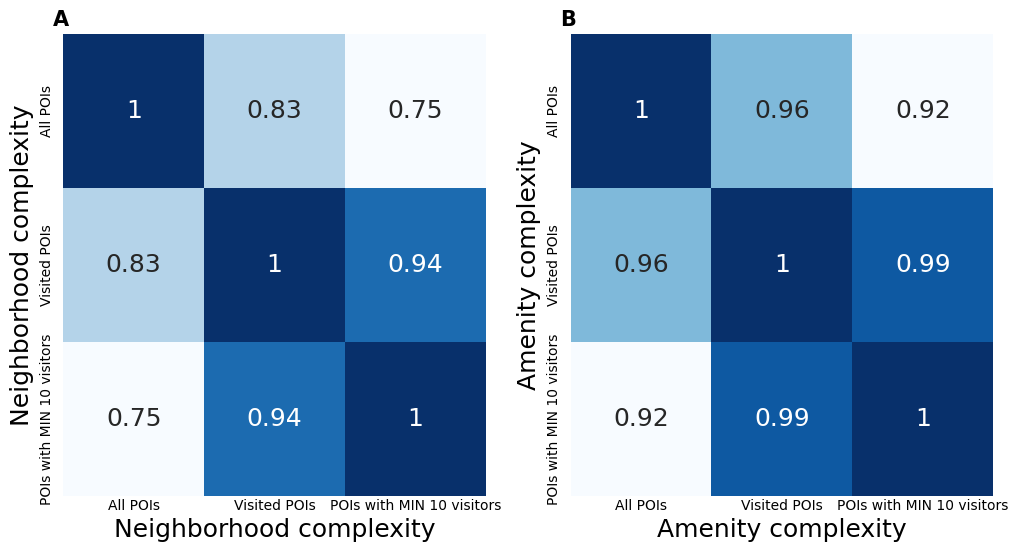}
\caption{Correlation tables for (A) neighborhood complexity and (B) amenity complexity values created from the neighborhood-amenity category matrix where all POIs, visited POIs or POIs with at least 10 visitor devices during the observed 24 months are considered.}
\label{fig:si4_fig4}
\end{figure}

As an alternative specification, we constructed neighborhood and amenity complexity measures based only on POIs that attract visitors along the 24 month long observed period. Figure~\ref{fig:si4_fig3} illustrates the stability of RCA 0/1 values in case we only consider POIs around Budapest visited by at least 1 or at least 10 devices during the 24 months. The figure suggests that 94\% of all neighborhood-amenity pairs are classified by the same RCA value even in case we consider the attraction of POIs at the beginning of the complexity measure generation. Changing from 1 to 10 minimum visitor devices to POIs does not change the classification results. Figure~\ref{fig:si4_fig4} presents strong correlations between neighborhood and amenity complexity values based on the different settings. Table~\ref{table:si4_table3} and Table~\ref{table:si4_table4} presents that our main results remain the same even in case we build our complexity measures on POIs that actually attract visitors.

\begin{table}[ht] \centering 
  \caption{Controlled correlations between the diversity of visitors and the complexity of neighborhoods (in 2020 February) in case we consider POIs with visitors for the neighborhood complexity measurement.} 
  \label{table:si4_table3}
\begin{tabular}{@{\extracolsep{5pt}}lccc} 
\hline \\[-1.8ex] 
\\[-1.8ex] & \multicolumn{3}{c}{Coefficient of variation} \\ 
\\[-1.8ex] & (1) & (2) & (3)\\ 
\hline \\[-1.8ex] 
 Neighborhood complexity & 0.125$^{***}$ &  &  \\ 
  & (0.036) &  &  \\ 
  & & & \\ 
 Neighborhood complexity visited POIs &  & 0.102$^{***}$ &  \\ 
  &  & (0.039) &  \\ 
  & & & \\ 
 Neighborhood complexity POIs with min 10 visitors &  &  & 0.101$^{**}$ \\ 
  &  &  & (0.041) \\ 
  & & & \\ 
 Centrality of location & 0.089$^{***}$ & 0.107$^{***}$ & 0.107$^{***}$ \\ 
  & (0.028) & (0.028) & (0.028) \\ 
  & & & \\ 
 Population (log) & $-$0.050$^{**}$ & $-$0.044$^{*}$ & $-$0.039 \\ 
  & (0.022) & (0.023) & (0.024) \\ 
  & & & \\ 
 Nr visitors (log) & 0.039 & 0.026 & 0.029 \\ 
  & (0.029) & (0.030) & (0.030) \\ 
  & & & \\ 
 Nr POIs (log) & 0.006 & 0.006 & $-$0.007 \\ 
  & (0.023) & (0.023) & (0.023) \\ 
  & & & \\ 
 Constant & 0.363$^{***}$ & 0.380$^{***}$ & 0.379$^{***}$ \\ 
  & (0.057) & (0.059) & (0.059) \\ 
  & & & \\ 
\hline \\[-1.8ex] 
Observations & 186 & 182 & 181 \\ 
R$^{2}$ & 0.303 & 0.281 & 0.267 \\ 
Adjusted R$^{2}$ & 0.284 & 0.261 & 0.247 \\ 
\hline \\[-1.8ex] 
\textit{Note:}  & \multicolumn{3}{r}{$^{*}$p$<$0.1; $^{**}$p$<$0.05; $^{***}$p$<$0.01} \\ 
\end{tabular} 
\end{table}

\begin{table}[ht] \centering 
  \caption{Controlled correlations between the diversity of visitors and the complexity of amenities (in 2020 February) in case we consider POIs with visitors for the amenity complexity measurement.} 
  \label{table:si4_table4}  
\begin{tabular}{@{\extracolsep{5pt}}lccc} 
\hline \\[-1.8ex] 
\\[-1.8ex] & \multicolumn{3}{c}{Coefficient of variation} \\ 
\\[-1.8ex] & (1) & (2) & (3)\\ 
\hline \\[-1.8ex] 
 Amenity complexity & 0.069$^{***}$ &  &  \\ 
  & (0.016) &  &  \\ 
  & & & \\ 
 Amenity complexity visited POIs &  & 0.075$^{***}$ &  \\ 
  &  & (0.017) &  \\ 
  & & & \\ 
 Amenity complexity POIs with min 10 visitors &  &  & 0.075$^{***}$ \\ 
  &  &  & (0.018) \\ 
  & & & \\ 
 Centrality of location & 0.116$^{***}$ & 0.115$^{***}$ & 0.115$^{***}$ \\ 
  & (0.011) & (0.011) & (0.011) \\ 
  & & & \\ 
 Population (log) & $-$0.006 & 0.001 & 0.004 \\ 
  & (0.007) & (0.007) & (0.008) \\ 
  & & & \\ 
 Nr visitors (log) & 0.069$^{***}$ & 0.069$^{***}$ & 0.070$^{***}$ \\ 
  & (0.008) & (0.007) & (0.007) \\ 
  & & & \\ 
 Nr POIs (log) & 0.172$^{***}$ & 0.154$^{***}$ & 0.148$^{***}$ \\ 
  & (0.030) & (0.030) & (0.031) \\ 
  & & & \\ 
\hline \\[-1.8ex] 
Observations & 2,742 & 2,742 & 2,742 \\ 
R$^{2}$ & 0.108 & 0.108 & 0.108 \\ 
Adjusted R$^{2}$ & 0.106 & 0.106 & 0.106 \\ 
\hline 
\textit{Note:}  & \multicolumn{3}{r}{$^{*}$p$<$0.1; $^{**}$p$<$0.05; $^{***}$p$<$0.01} \\ 
\end{tabular} 
\end{table}

\clearpage
\section*{S5 Similarity of amenities based on their co-location}

\begin{figure}[!ht]
\centering
\includegraphics[width=0.85\textwidth]{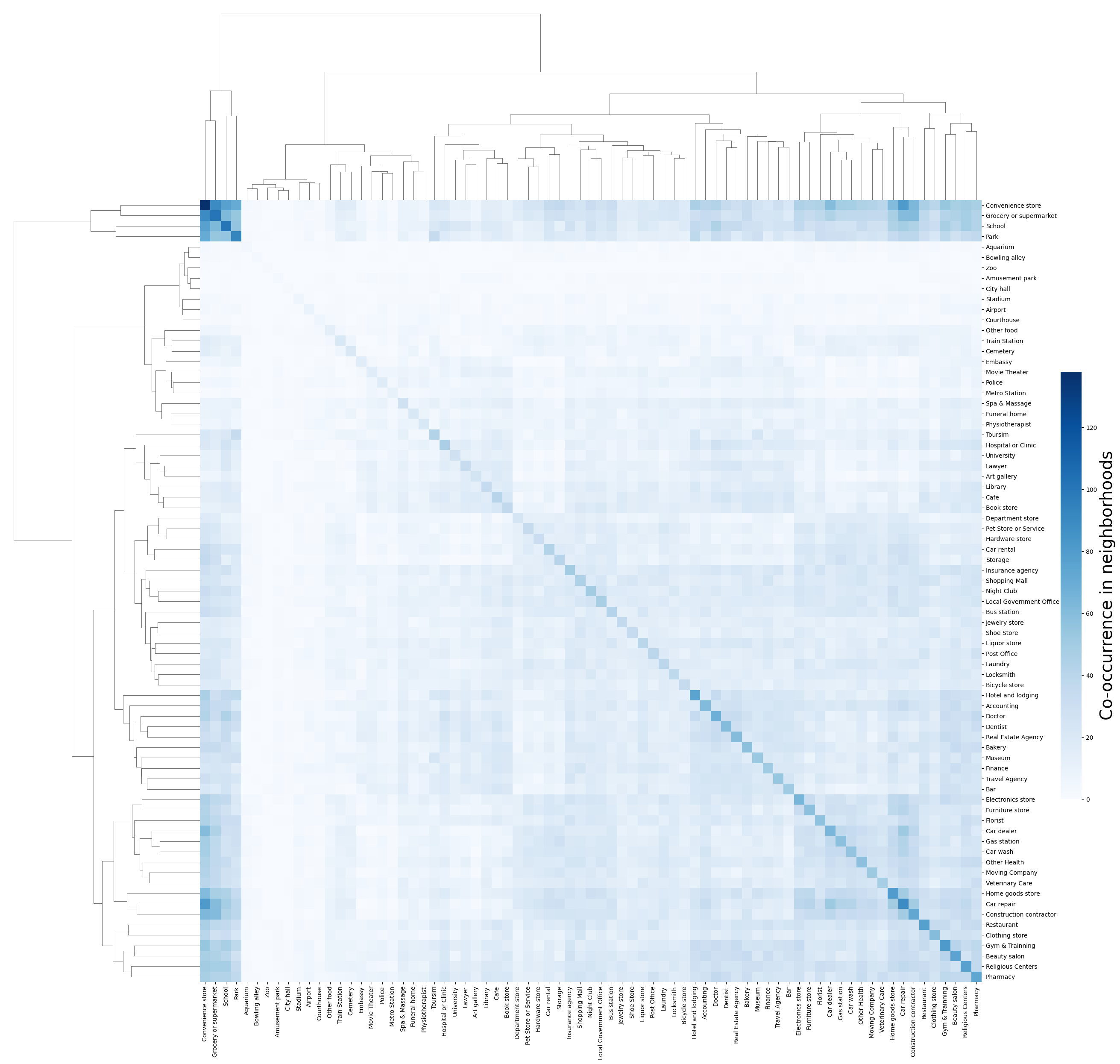}
\caption{Clustered version of the amenity-amenity similarity matrix $M_{aa'}$. Colors of the matrix indicate the number of neighborhoods we observe two amenity categories with RCA value above 1.}
\label{fig:si5_fig1}
\end{figure}

\begin{figure}[!ht]
\centering
\includegraphics[width=0.85\textwidth]{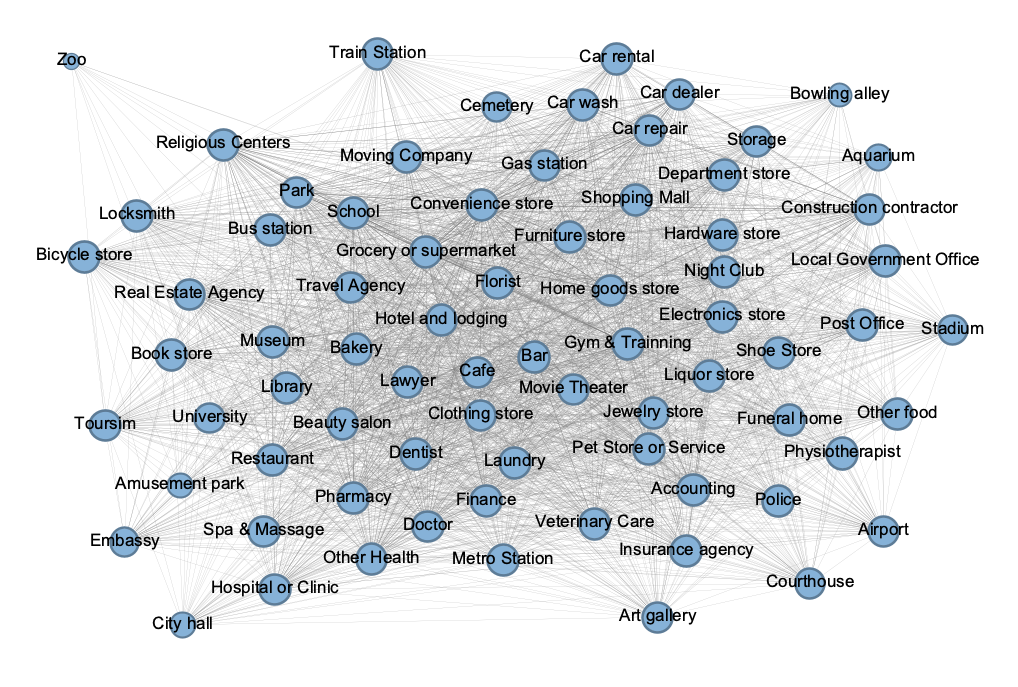}
\caption{Network representation of the amenity-amenity similarity matrix $M_{aa'}$. Edges represent non-normalized values of co-concentration in neighborhoods.} 
\label{fig:si5_fig2}
\end{figure}

Figure~\ref{fig:si5_fig1} presents an alternative, clustered version of the $M_{aa'}$ amenity-amenity similarity matrix, which we use to construct our amenity complexity measure. The strongest stand-alone clusters consist of ubiquitous amenities such as Grocery or supermarket and Convenience store, which co-occur in many neighborhoods with a concentration of $RCA>=1$. A normalized version of this matrix is often represented as a network, similarly to the product space or the knowledge space \cite{hidalgo2021review}. Figure~\ref{fig:si5_fig2} illustrates our amenity-amenity similarity matrix as a network with non-normalized edge weights.

\clearpage
\section*{S6 Amenity complexity rankings}

\begin{figure}[htb]
\centering
\includegraphics[width=0.65\textwidth]{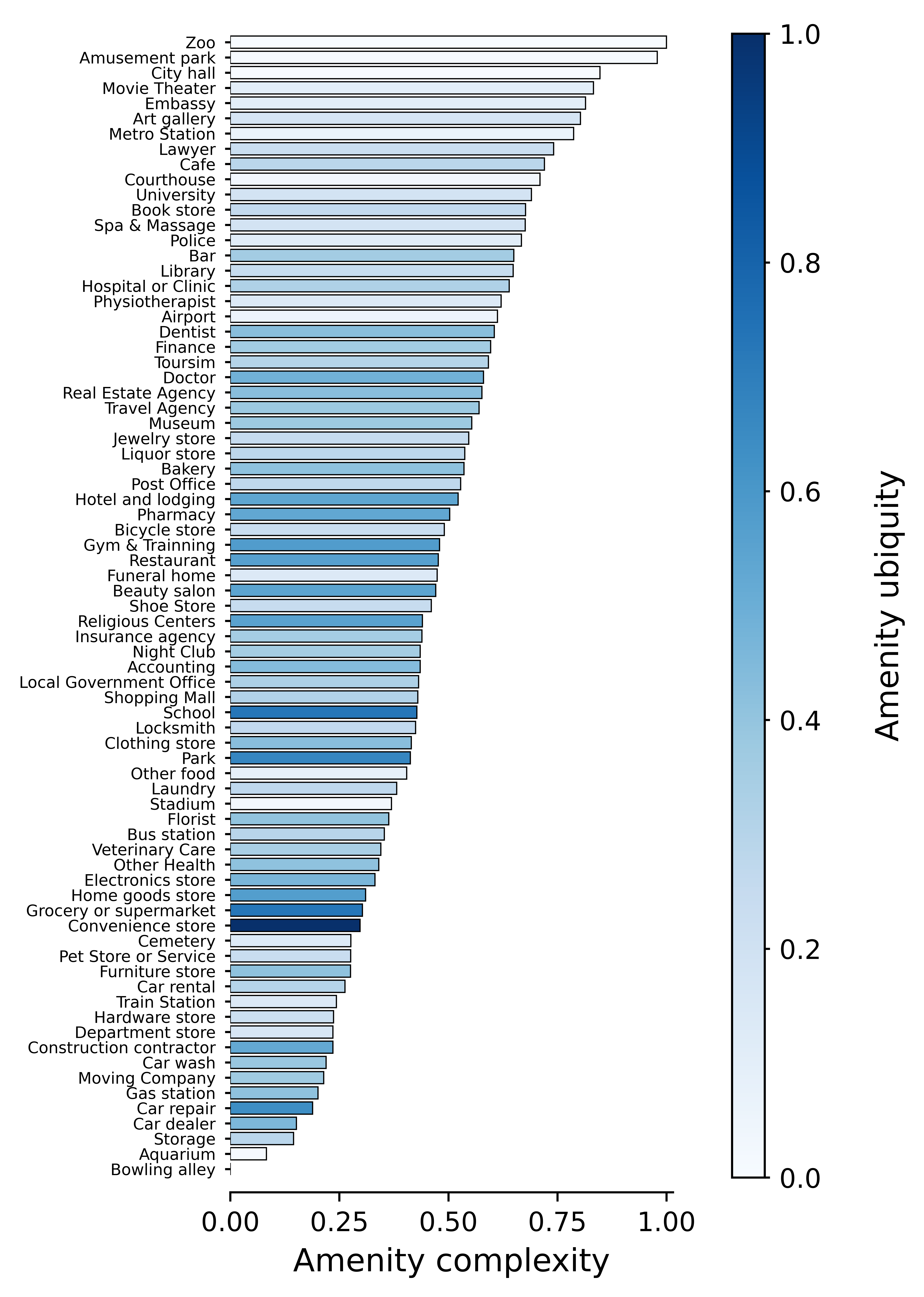}
\caption{Amenity categories ranked by their amenity complexity value. Categories are colored by their ubiquity across neighborhoods. Complexity and ubiquity values are normalized to 0-1 scale for visualization purposes.}
\label{fig:si6_fig1}
\end{figure}

\begin{figure}[htb]
\centering
\includegraphics[width=0.9\textwidth]{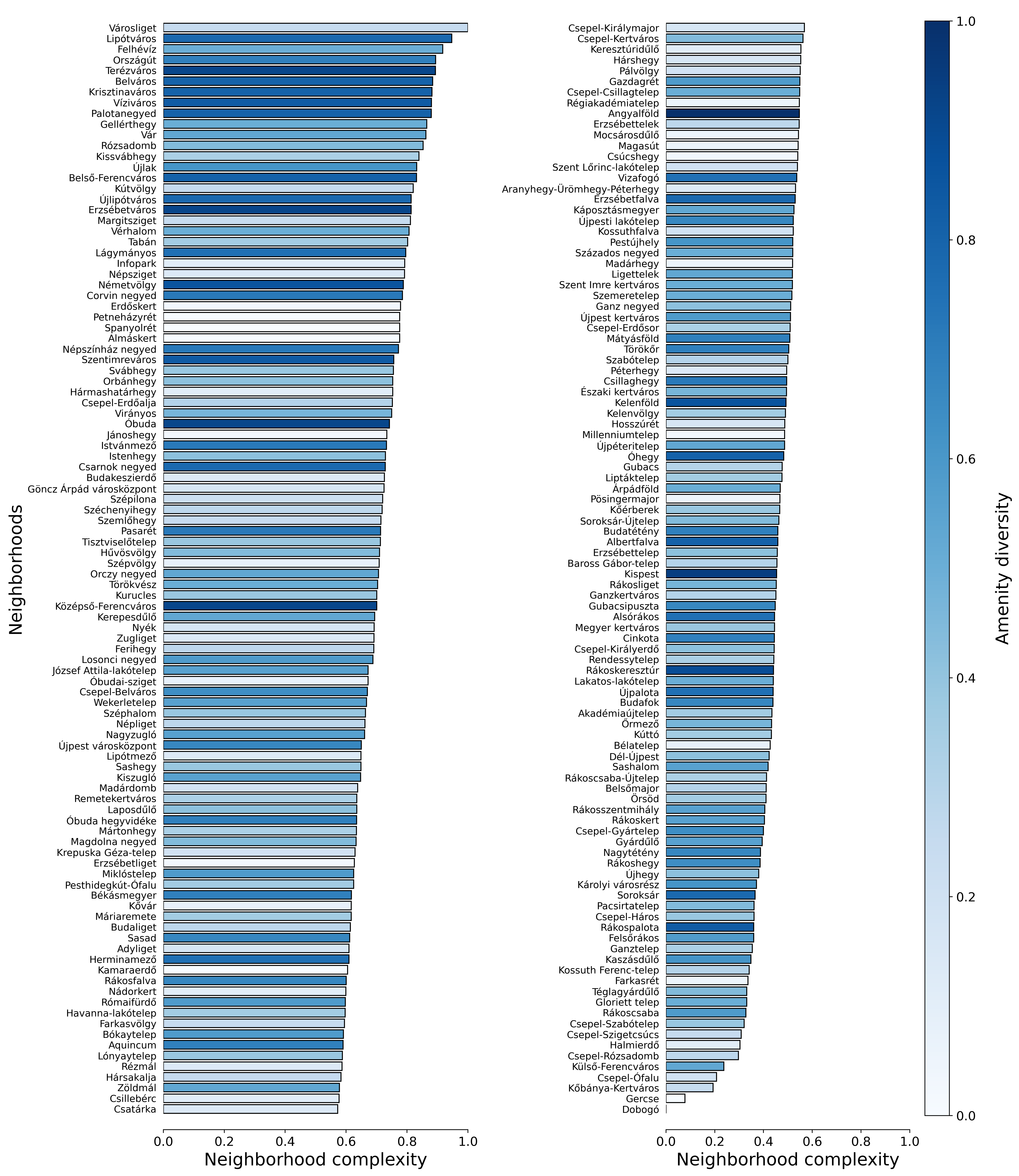}
\caption{Neighborhoods of Budapest ranked by their amenity complexity value. Neighborhoods are colored by the diversity of their amenities. Complexity and diversity values are normalized to 0-1 scale for visualization purposes.}
\label{fig:si6_fig2}
\end{figure}

Ranking of all amenity categories (Figure \ref{fig:si6_fig1}) and all neighborhoods by their amenity complexity values (Figure \ref{fig:si6_fig2}).

\clearpage

\clearpage
\section*{S7 Centrality of locations in Budapest, Hungary}

\begin{figure}[htb]
\centering
\includegraphics[width=0.9\textwidth]{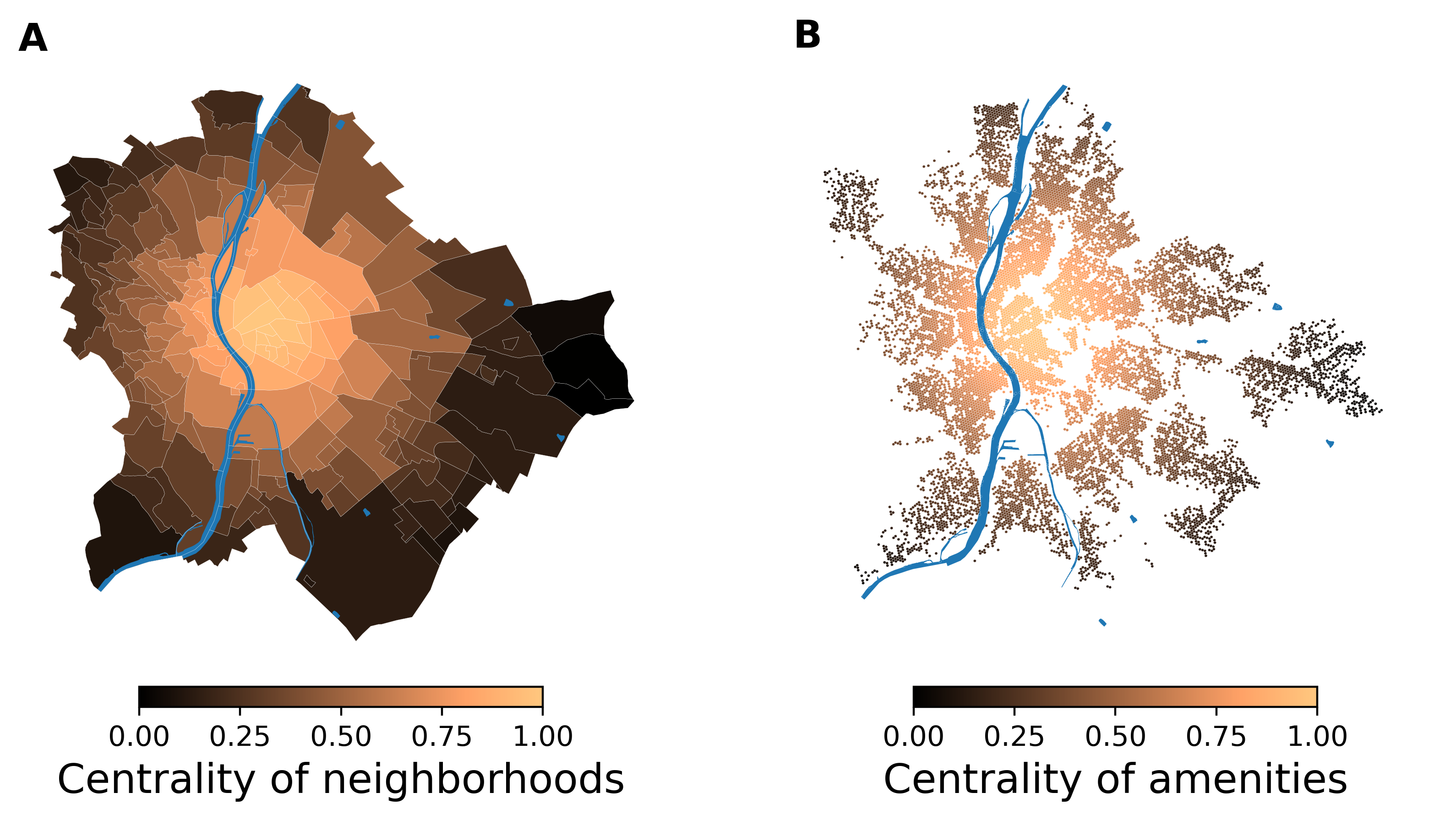}
\caption{Centrality of locations (A) Centrality of neighborhoods on the map of Budapest, Hungary. Centrality is measured as the inverse of the average distance to reach a neighborhood from all the census tracts of Budapest. (B) Centrality of amenities on the map of Budapest, Hungary. Centrality is measured by the inverse average distance to reach an amenity from all the census tracts of Budapest. Areas in blue indicate natural water sources.}
\label{fig:si7_fig1}
\end{figure}

\begin{figure}[htb]
\centering
\includegraphics[width=0.9\textwidth]{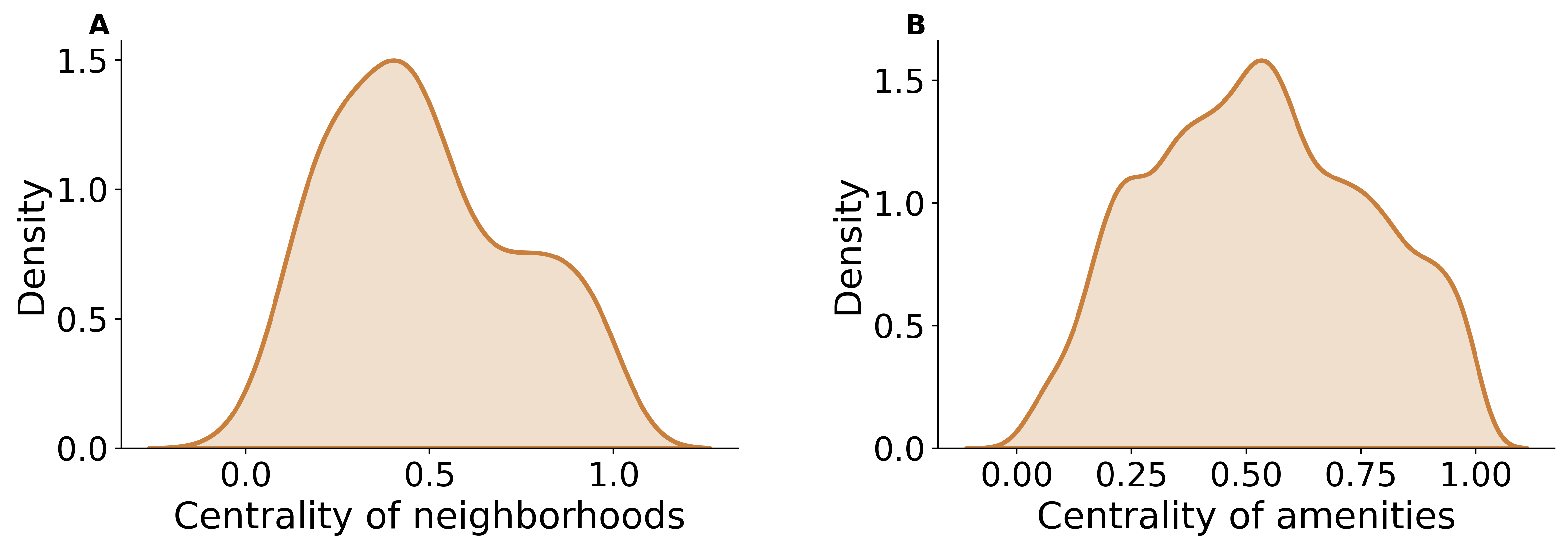}
\caption{Distribution of centrality measures for (A) neighborhoods and (B) amenities. Centrality of locations is defined as the inverse of the average distance to reach the location from all the census tracts of Budapest. Areas in blue indicate natural water sources.}
\label{fig:si7_fig2}
\end{figure}

To identify central locations within the city of Budapest, we measure the average distance of locations from each census tract of the city. More precisely, for neighborhoods we measure the Euclidean distance between the centroid of the focal neighborhood and the centroid of all census tracts and calculate the average distance to reach the neighborhood from any census tracts of the city. For actual amenities, we apply the same procedure and calculate the average distance between the centroid of the focal H3 hexagon and the centroid of all census tracts. For both neighborhoods and amenities, we use the inverse of this average distance (log) normalized to 0-1 scale. Figure~\ref{fig:si7_fig1} illustrates the resulted indicators on the map of Budapest, where higher values are associated with locations around the historic inner city of Pest. Figure~\ref{fig:si7_fig2} presents the distribution of the indicator for neighborhoods and amenities.

We choose this solution to quantify the centrality of locations for several reasons. First, Budapest has a convex shape and it is a clearly monocentric city. The application of the Urban Centrality Index (UCI) proposed by \textcite{pereira2013urbancentrality} to the population distribution between Budapest neighborhoods results in Proximity Index (PI) of 0.999 and UCI of 0.386, indicating a highly centralized structure.
Second, we take advantage of the census tract structure. Census tracts are relatively homogeneous in terms of population, but very heterogeneous in terms of area. More specifically, census tracts tend to be smaller around the city center and larger on the outskirts. The densely populated inner city, which concentrates a large proportion of census tracts results in many neighborhood-census tract pairs (and amenity-census tract pairs) with shorter distances. As a result, the average value of distances will be high for densely populated central locations.
Third, this method allows assigning a unique centrality value to each neighborhood and each H3 hexagon based on their relationship to the common spatial unit of the census tracts.
Finally, our metric is universal in the sense that it can be applied to any city of similar shape and structure to identify the centrality of locations, and is less city-specific than measuring the distance to an arbitrary central location chosen based on local knowledge.

Table~\ref{table:si7_table1} shows how different measures on centrality of locations affect the relationship between socio-economic diversity of visitors. The regressions are based on the example month of February 2020, exactly like our models in the main text. Model (1) is identical with our preferred model from the main text. Model (2) tests the significance of the distance from the center of mass in Budapest. Model (3) uses distance from the zero kilometer stone as an alternative centrality measure. The zero kilometer stone (or landmark zero sculpture) at Clark Ádám tér is the center of Hungary's main road network, and road network distances are calculated from here. Model (4) introduces the distance from the center of Budapest defined by Deák Ferenc tér. This square is a public transport hotspot in the heart of the city center.
For model (5) we compute the Local Indicator of Spatial Association (LISA) using the local Moran's I for the number of POIs per neighborhood \cite{anselin1995lisa}. More precisely, we find the largest, statistically significant spatial cluster of neighborhoods in which the number of POIs and the average number of POIs in its surroundings is more similar than would be expected by pure chance \citep{reyetal2023geodatasci}. The distance from LISA center variable measures the distance of neighborhoods from the center of the largest statistically significant spatial cluster. The resulting center is near the Nyugati Railway Station, one of the largest railway stations in the city. All the above variables capturing urban centrality are log transferred. As all variables for urban centrality are significant and have the expected sign (except the model using center of mass), Table~\ref{table:si7_table1} clearly illustrates that urban centrality is related to socio-economic mixing.

\begin{table}[ht] \centering 
  \caption{Controlled correlations between the socio-economic diversity of visitors and different measures on central location of neighborhoods} 
  \label{table:si7_table1} 
\begin{tabular}{@{\extracolsep{5pt}}lccccc} 
\hline \\[-1.8ex] 
\\[-1.8ex] & \multicolumn{5}{c}{Coefficient of variation} \\ 
\\[-1.8ex] & (1) & (2) & (3) & (4) & (5)\\ 
\hline \\[-1.8ex] 
 Centrality of location & 0.130$^{***}$ &  &  &  &  \\ 
  & (0.027) &  &  &  &  \\ 
  & & & & & \\ 
 Distance from center of mass &  & $-$0.015 &  &  &  \\ 
  &  & (0.025) &  &  &  \\ 
  & & & & & \\ 
 Distance from zero km stone &  &  & $-$0.151$^{***}$ &  &  \\ 
  &  &  & (0.019) &  &  \\ 
  & & & & & \\ 
 Distance from Deak ter &  &  &  & $-$0.151$^{***}$ &  \\ 
  &  &  &  & (0.022) &  \\ 
  & & & & & \\ 
 Distance from LISA center &  &  &  &  & $-$0.163$^{***}$ \\ 
  &  &  &  &  & (0.023) \\ 
  & & & & & \\ 
 Population (log) & $-$0.053$^{**}$ & $-$0.096$^{***}$ & $-$0.040$^{**}$ & $-$0.038$^{*}$ & $-$0.038$^{*}$ \\ 
  & (0.022) & (0.023) & (0.020) & (0.021) & (0.021) \\ 
  & & & & & \\ 
 Nr visitors (log) & 0.047 & 0.099$^{***}$ & 0.036 & 0.034 & 0.024 \\ 
  & (0.029) & (0.029) & (0.026) & (0.027) & (0.027) \\ 
  & & & & & \\ 
 Nr POIs (log) & $-$0.001 & 0.005 & 0.004 & $-$0.0001 & 0.004 \\ 
  & (0.023) & (0.025) & (0.021) & (0.022) & (0.022) \\ 
  & & & & & \\ 
 Constant & 0.424$^{***}$ & 0.536$^{***}$ & 0.576$^{***}$ & 0.583$^{***}$ & 0.606$^{***}$ \\ 
  & (0.056) & (0.056) & (0.048) & (0.049) & (0.049) \\ 
  & & & & & \\ 
\hline \\[-1.8ex] 
Observations & 186 & 186 & 186 & 186 & 186 \\ 
R$^{2}$ & 0.257 & 0.162 & 0.376 & 0.335 & 0.350 \\ 
Adjusted R$^{2}$ & 0.241 & 0.144 & 0.363 & 0.321 & 0.336 \\ 
\hline \\[-1.8ex] 
\textit{Note:}  & \multicolumn{5}{r}{$^{*}$p$<$0.1; $^{**}$p$<$0.05; $^{***}$p$<$0.01} \\ 
\end{tabular} 
\end{table}

Table~\ref{table:si7_table2} presents how different measures on centrality of locations affect the relationship between socio-economic diversity of visitors and neighborhood complexity. Model (1) is used as a baseline, in which we do not control for the centrality of neighborhoods at all. From model (2) to model (6), we present one by one the measures of location centrality presented earlier. The table shows mixed results. While the distance from the center of mass in Budapest does not influence the diversity of visitors, the distance from infrastructure hubs such as the zero kilometer stone, Deak tér or Nyugati railway station, which is the result of the LISA calculation, have high explanatory power.

\begin{table}[ht] \centering 
  \caption{Controlled correlations between the socio-economic diversity of visitors, complexity of neighborhoods and alternative measures on central location} 
  \label{table:si7_table2} 
\begin{tabular}{@{\extracolsep{1pt}}lcccccc} 
\hline \\[-1.8ex] 
\\[-1.8ex] & \multicolumn{6}{c}{Coefficient of variation} \\ 
\\[-1.8ex] & (1) & (2) & (3) & (4) & (5) & (6)\\ 
\hline \\[-1.8ex] 
 Neighborhood complexity & 0.172$^{***}$ & 0.125$^{***}$ & 0.172$^{***}$ & 0.028 & 0.056 & 0.042 \\ 
  & (0.034) & (0.036) & (0.034) & (0.040) & (0.041) & (0.041) \\ 
  & & & & & & \\ 
 Centrality of location &  & 0.089$^{***}$ &  &  &  &  \\ 
  &  & (0.028) &  &  &  &  \\ 
  & & & & & & \\ 
 Distance from center of mass &  &  & $-$0.005 &  &  &  \\ 
  &  &  & (0.024) &  &  &  \\ 
  & & & & & & \\ 
 Distance from zero km stone &  &  &  & $-$0.140$^{***}$ &  &  \\ 
  &  &  &  & (0.025) &  &  \\ 
  & & & & & & \\ 
 Distance from Deak ter &  &  &  &  & $-$0.127$^{***}$ &  \\ 
  &  &  &  &  & (0.028) &  \\ 
  & & & & & & \\ 
 Distance from LISA center &  &  &  &  &  & $-$0.144$^{***}$ \\ 
  &  &  &  &  &  & (0.029) \\ 
  & & & & & & \\ 
 Population (log) & $-$0.075$^{***}$ & $-$0.050$^{**}$ & $-$0.074$^{***}$ & $-$0.040$^{**}$ & $-$0.040$^{*}$ & $-$0.039$^{*}$ \\ 
  & (0.021) & (0.022) & (0.022) & (0.020) & (0.021) & (0.021) \\ 
  & & & & & & \\ 
 Nr visitors (log) & 0.067$^{**}$ & 0.039 & 0.067$^{**}$ & 0.035 & 0.033 & 0.025 \\ 
  & (0.028) & (0.029) & (0.028) & (0.026) & (0.027) & (0.027) \\ 
  & & & & & & \\ 
 Nr POIs (log) & 0.013 & 0.006 & 0.012 & 0.005 & 0.003 & 0.006 \\ 
  & (0.023) & (0.023) & (0.023) & (0.021) & (0.022) & (0.022) \\ 
  & & & & & & \\ 
 Constant & 0.399$^{***}$ & 0.363$^{***}$ & 0.402$^{***}$ & 0.552$^{***}$ & 0.533$^{***}$ & 0.566$^{***}$ \\ 
  & (0.057) & (0.057) & (0.059) & (0.059) & (0.062) & (0.063) \\ 
  & & & & & & \\ 
\hline \\[-1.8ex] 
Observations & 186 & 186 & 186 & 186 & 186 & 186 \\ 
R$^{2}$ & 0.265 & 0.303 & 0.266 & 0.378 & 0.342 & 0.354 \\ 
Adjusted R$^{2}$ & 0.249 & 0.284 & 0.245 & 0.361 & 0.324 & 0.336 \\ 
\hline \\[-1.8ex] 
\textit{Note:}  & \multicolumn{6}{r}{$^{*}$p$<$0.1; $^{**}$p$<$0.05; $^{***}$p$<$0.01} \\ 
\end{tabular} 
\end{table} 

Table~\ref{table:si7_table3} and Table~\ref{table:si7_table4} apply the same logic and setting to amenities. Table~\ref{table:si7_table3} shows that measures of centrality of location work the same for amenities as for neighborhoods, and that amenities in more central locations are visited by socio-economically more diverse people. Table~\ref{table:si7_table4} presents models in which we also control for amenity complexity. The results are mixed, and unlike at the neighborhood level, amenity complexity is significant even after controlling for distance from Deák tér or for the distance from the LISA center (Nyugati railway station area). These regressions suggest that in addition to the physical centrality of locations, centrality within the transport infrastructure may also influence socio-economic mixing in Budapest, Hungary. This would require further evaluation, as the zero kilometer stone, for example, which has the largest explanatory power in both settings, is only a symbolic center of the Hungarian road network. It does not serve as a meeting point for the local population, nor can it be seen as a melting pot of society. However, a deeper assessment is beyond the scope of this research.

\begin{table}[ht] \centering 
  \caption{Controlled correlations between the socio-economic diversity of visitors and different measures on central location of amenities} 
  \label{table:si7_table3} 
\begin{tabular}{@{\extracolsep{5pt}}lccccc} 
\hline \\[-1.8ex] 
\\[-1.8ex] & \multicolumn{5}{c}{Coefficient of variation} \\ 
\\[-1.8ex] & (1) & (2) & (3) & (4) & (5)\\ 
\hline \\[-1.8ex] 
 Centrality of location & 0.126$^{***}$ &  &  &  &  \\ 
  & (0.011) &  &  &  &  \\ 
  & & & & & \\ 
 Distance from center of mass &  & $-$0.011 &  &  &  \\ 
  &  & (0.011) &  &  &  \\ 
  & & & & & \\ 
 Distance from zero km stone &  &  & $-$0.132$^{***}$ &  &  \\ 
  &  &  & (0.011) &  &  \\ 
  & & & & & \\ 
 Distance from Deak ter &  &  &  & $-$0.115$^{***}$ &  \\ 
  &  &  &  & (0.010) &  \\ 
  & & & & & \\ 
 Distance from LISA center &  &  &  &  & $-$0.109$^{***}$ \\ 
  &  &  &  &  & (0.012) \\ 
  & & & & & \\ 
 Nr POIs in category (log) & $-$0.013 & $-$0.021 & $-$0.006 & $-$0.006 & $-$0.010 \\ 
  & (0.009) & (0.013) & (0.007) & (0.007) & (0.008) \\ 
  & & & & & \\ 
 Nr visitors (log) & 0.073$^{***}$ & 0.096$^{***}$ & 0.053$^{***}$ & 0.047$^{***}$ & 0.056$^{***}$ \\ 
  & (0.007) & (0.009) & (0.006) & (0.007) & (0.008) \\ 
  & & & & & \\ 
 Constant & 0.220$^{***}$ & 0.313$^{***}$ & 0.397$^{***}$ & 0.388$^{***}$ & 0.387$^{***}$ \\ 
  & (0.033) & (0.051) & (0.030) & (0.032) & (0.034) \\ 
  & & & & & \\ 
\hline \\[-1.8ex] 
Observations & 2,742 & 2,742 & 2,742 & 2,742 & 2,742 \\ 
R$^{2}$ & 0.102 & 0.052 & 0.167 & 0.143 & 0.138 \\ 
Adjusted R$^{2}$ & 0.101 & 0.051 & 0.166 & 0.142 & 0.137 \\ 
\hline \\[-1.8ex] 
\textit{Note:}  & \multicolumn{5}{r}{$^{*}$p$<$0.1; $^{**}$p$<$0.05; $^{***}$p$<$0.01} \\ 
\end{tabular} 
\end{table}

\begin{table}[!htbp] \centering 
  \caption{Controlled correlations between the socio-economic diversity of visitors, amenity complexity and alternative measures on central location} 
  \label{table:si7_table4}
\begin{tabular}{@{\extracolsep{1pt}}lcccccc} 
\hline \\[-1.8ex] 
\\[-1.8ex] & \multicolumn{6}{c}{Coefficient of variation} \\ 
\\[-1.8ex] & (1) & (2) & (3) & (4) & (5) & (6)\\ 
\hline \\[-1.8ex] 
 Amenity complexity & 0.113$^{***}$ & 0.069$^{***}$ & 0.112$^{***}$ & 0.019 & 0.031$^{**}$ & 0.046$^{***}$ \\ 
  & (0.021) & (0.016) & (0.021) & (0.015) & (0.015) & (0.014) \\ 
  & & & & & & \\ 
 Centrality of location &  & 0.116$^{***}$ &  &  &  &  \\ 
  &  & (0.011) &  &  &  &  \\ 
  & & & & & & \\ 
 Distance from center of mass &  &  & $-$0.003 &  &  &  \\ 
  &  &  & (0.011) &  &  &  \\ 
  & & & & & & \\ 
 Distance from zero km stone &  &  &  & $-$0.129$^{***}$ &  &  \\ 
  &  &  &  & (0.011) &  &  \\ 
  & & & & & & \\ 
 Distance from Deak ter &  &  &  &  & $-$0.111$^{***}$ &  \\ 
  &  &  &  &  & (0.009) &  \\ 
  & & & & & & \\ 
 Distance from LISA center &  &  &  &  &  & $-$0.104$^{***}$ \\ 
  &  &  &  &  &  & (0.012) \\ 
  & & & & & & \\ 
 Nr POIs in category (log) & $-$0.008 & $-$0.006 & $-$0.008 & $-$0.004 & $-$0.003 & $-$0.006 \\ 
  & (0.008) & (0.007) & (0.008) & (0.006) & (0.007) & (0.007) \\ 
  & & & & & & \\ 
 Nr visitors (log) & 0.087$^{***}$ & 0.069$^{***}$ & 0.086$^{***}$ & 0.052$^{***}$ & 0.046$^{***}$ & 0.054$^{***}$ \\ 
  & (0.010) & (0.008) & (0.010) & (0.006) & (0.007) & (0.008) \\ 
  & & & & & & \\ 
 Constant & 0.216$^{***}$ & 0.172$^{***}$ & 0.219$^{***}$ & 0.381$^{***}$ & 0.361$^{***}$ & 0.347$^{***}$ \\ 
  & (0.034) & (0.030) & (0.035) & (0.032) & (0.031) & (0.030) \\ 
  & & & & & & \\ 
\hline \\[-1.8ex] 
Observations & 2,742 & 2,742 & 2,742 & 2,742 & 2,742 & 2,742 \\ 
R$^{2}$ & 0.067 & 0.108 & 0.067 & 0.167 & 0.144 & 0.140 \\ 
Adjusted R$^{2}$ & 0.066 & 0.106 & 0.066 & 0.166 & 0.143 & 0.139 \\ 
\hline \\[-1.8ex] 
\textit{Note:}  & \multicolumn{6}{r}{$^{*}$p$<$0.1; $^{**}$p$<$0.05; $^{***}$p$<$0.01} \\ 
\end{tabular} 
\end{table}

\clearpage
\section*{S8 Dominant amenity categories of H3 hexagons}

\begin{figure}[H]
  \includegraphics[width=0.95\linewidth]{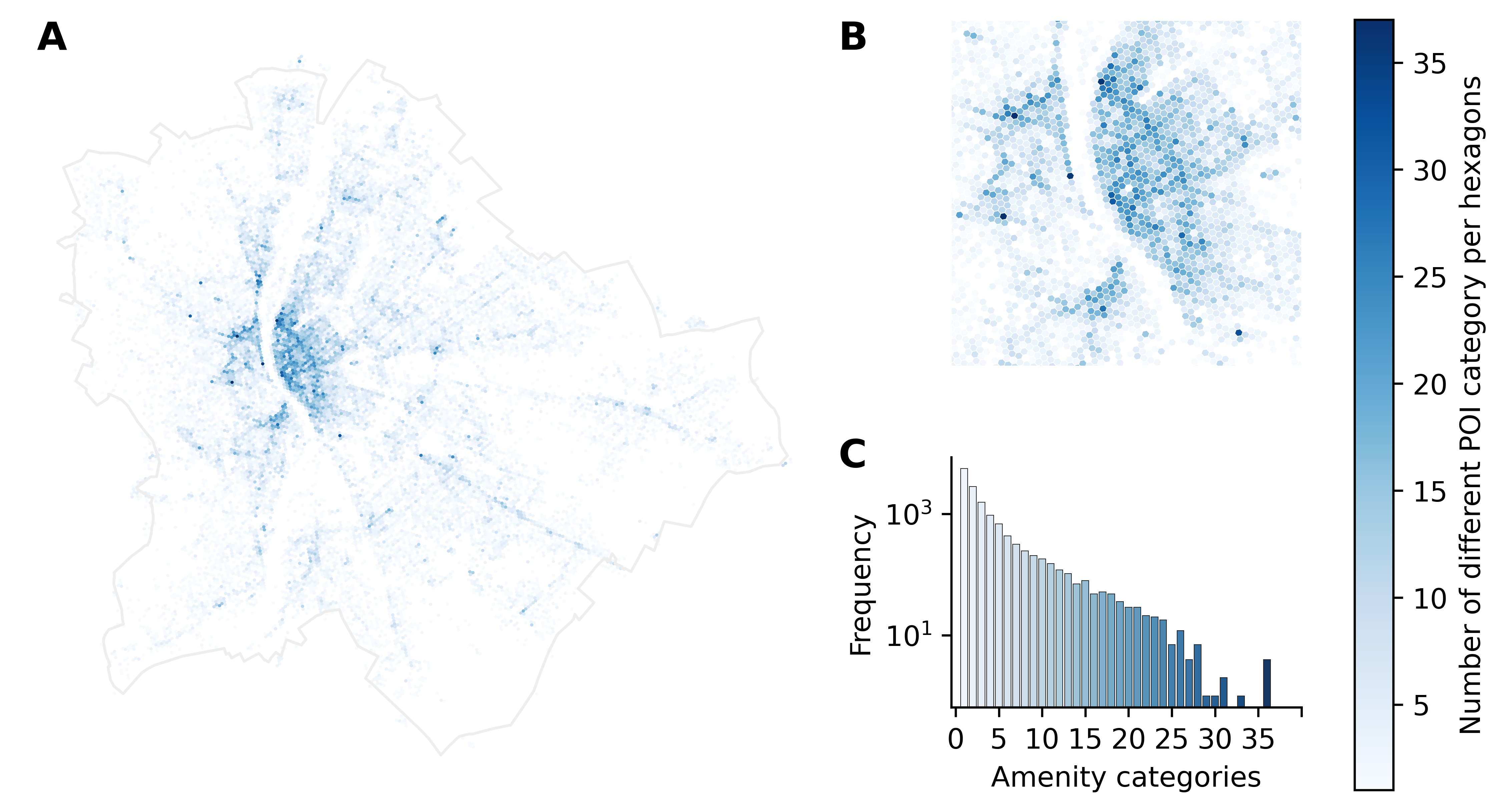}
  \caption{Number of different amenity categories per size 10 H3 hexagons on the map of Budapest (A) and on the map of the inner city (B). Distribution of different amenity categories per hexagons (C). The colorbar applies for all subplots.}
  \label{fig:si8_fig1}
\end{figure}

The second part of our empirical exercise (section 4.2 Diversity of visitors to complex amenities) connects visitors to actual amenities. This is done by mapping each point of interest (POI) from the Google Places API to a size 10 H3 hexagons (\cite{h3}). These hexagons are on average 15.000 $m^{2}$ area, which is close to the buffer area of a point with a 70 meter radius. As we use the amenity complexity values calculated for amenity categories (see Equation (6) in the main text) to explain the diversity of visitors to size 10 H3 hexagons, we need to assign a single amenity category to each hexagon with POIs. Figure~\ref{fig:si8_fig1} illustrates that most hexagons only have amenities in a single amenity category, however, hexagons at dense, central locations often contain multiple POIs from different categories.

\begin{figure}[H]
    \includegraphics[width=0.95\linewidth]{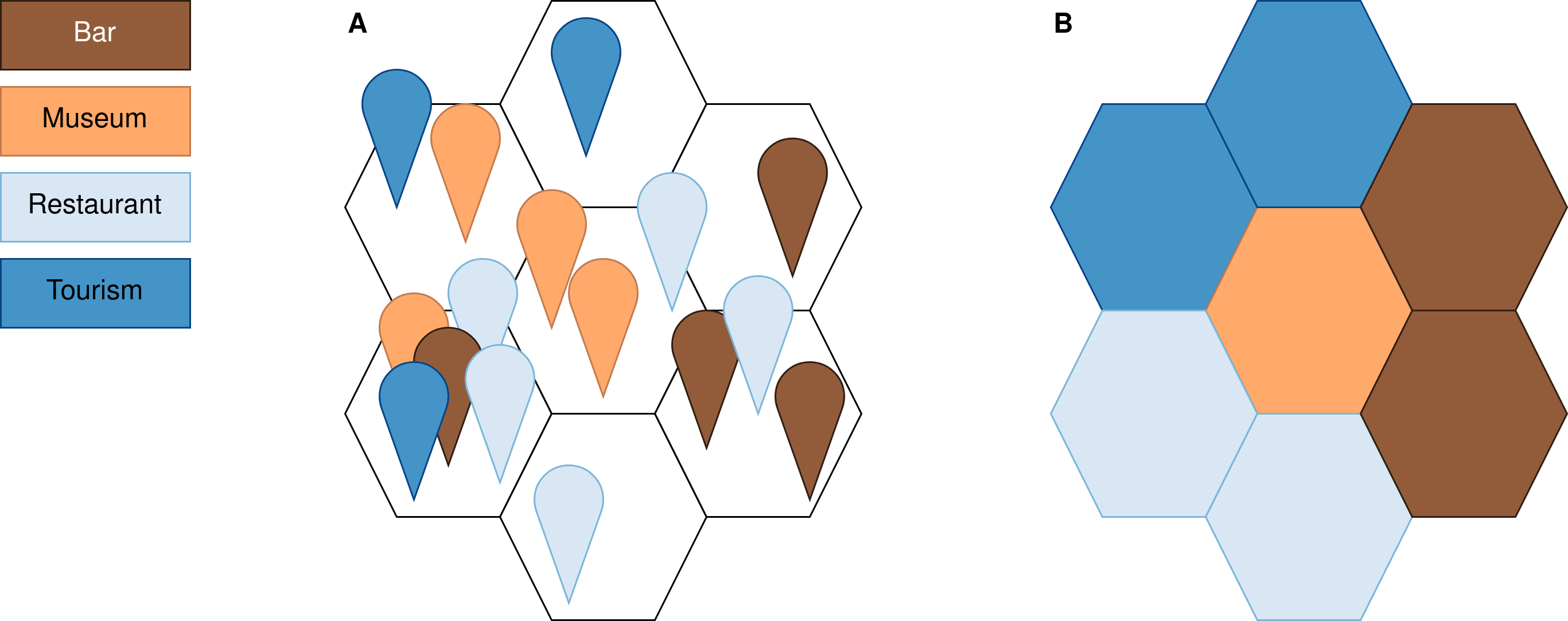}
  \caption{Illustration of the dominant amenity category selection. Different POIs over the hexagons (A) and the determined hexagon category (\textbf{b}).}
  \label{fig:si8_fig2}
\end{figure}

We assign a dominant amenity category for each hexagon based on the local frequency of POIs in amenity categories. Figure~\ref{fig:si8_fig2} visualizes our dominant category selection process. \num{35.15}\% of the hexagons have ambiguous amenity category dominance and in these cases we simply choose the first category listed. Table~\ref{table:si8_table1} illustrates in comparison to Table 2 of the main text that our key findings are the same in case we focus only on amenities in H3 hexagons with a single amenity category or in case we only consider amenities in H3 hexagons with an unambiguously dominant amenity category. 

\renewcommand{\arraystretch}{0.95}
\begin{table}[!ht] \centering 
  \caption{Controlled correlations between the diversity of visitors and the complexity of amenities in case we only consider H3 hexagons with (1) a single amenity category or with (2) an unambiguously dominant amenity category} 
  \label{table:si8_table1}
\begin{tabular}{@{\extracolsep{5pt}}lcc} 
\hline \\[-1.8ex] 
\\[-1.8ex] & \multicolumn{2}{c}{Coefficient of variation} \\ 
 & Single category & Dominant category \\ 
\\[-1.8ex] & (1) & (2)\\ 
\hline \\[-1.8ex] 
 Amenity complexity & 0.153$^{**}$ & 0.080$^{***}$ \\ 
  & (0.075) & (0.021) \\ 
  & & \\ 
 Centrality of location & 0.061 & 0.127$^{***}$ \\ 
  & (0.059) & (0.012) \\ 
  & & \\ 
 Nr POIs in category (log) & $-$0.032 & $-$0.010 \\ 
  & (0.024) & (0.008) \\ 
  & & \\ 
 Nr visitors (log) & 0.144$^{***}$ & 0.063$^{***}$ \\ 
  & (0.052) & (0.008) \\ 
  & & \\ 
 Constant & 0.154$^{*}$ & 0.181$^{***}$ \\ 
  & (0.080) & (0.034) \\ 
  & & \\ 
\hline \\[-1.8ex] 
Observations & 151 & 1,849 \\ 
R$^{2}$ & 0.091 & 0.127 \\ 
Adjusted R$^{2}$ & 0.066 & 0.125 \\ 
\hline \\[-1.8ex] 
\textit{Note:}  & \multicolumn{2}{r}{$^{*}$p$<$0.1; $^{**}$p$<$0.05; $^{***}$p$<$0.01} \\ 
\end{tabular} 
\end{table}

\clearpage
\section*{S9 Alternative measures on the diversity of visitors to neighborhoods and amenities}

Table~\ref{table:si9_table1} supports the findings presented in Table 1 of the main text. It uses the same model setting to illustrate the relationship between the diversity of visitors to neighborhoods, neighborhood complexity and urban centrality, but applies different measurements of the dependent variable. Besides the coefficient of variation, the Gini index and the Theil index are used to measure the diversity of visitors and all model version suggest similar connections.
Table~\ref{table:si9_table2} supplements the findings of Table 2 in the main text in a similar fashion at the level of amenities.

\renewcommand{\arraystretch}{0.95}
\begin{table}[!htbp] \centering 
  \caption{Controlled correlations between different measures on the diversity of visitors and neighborhood complexity} 
  \label{table:si9_table1} 
\begin{tabular}{@{\extracolsep{5pt}}lccc} 
\hline \\[-1.8ex] 
\\[-1.8ex] & Coefficient of variation & Gini & Theil \\ 
\\[-1.8ex] & (1) & (2) & (3)\\ 
\hline \\[-1.8ex] 
 Neighborhood complexity & 0.125$^{***}$ & 0.088$^{***}$ & 0.056$^{***}$ \\ 
  & (0.036) & (0.018) & (0.012) \\ 
  & & & \\ 
 Centrality of location & 0.089$^{***}$ & 0.032$^{**}$ & 0.019$^{*}$ \\ 
  & (0.028) & (0.014) & (0.010) \\ 
  & & & \\ 
 Population (log) & $-$0.050$^{**}$ & $-$0.026$^{**}$ & $-$0.016$^{**}$ \\ 
  & (0.022) & (0.011) & (0.007) \\ 
  & & & \\ 
 Nr visitors (log) & 0.039 & 0.018 & 0.011 \\ 
  & (0.029) & (0.014) & (0.010) \\ 
  & & & \\ 
 Nr POIs (log) & 0.006 & $-$0.002 & $-$0.002 \\ 
  & (0.023) & (0.011) & (0.008) \\ 
  & & & \\ 
 Constant & 0.363$^{***}$ & 0.198$^{***}$ & 0.068$^{***}$ \\ 
  & (0.057) & (0.029) & (0.019) \\ 
  & & & \\ 
\hline \\[-1.8ex] 
Observations & 186 & 186 & 186 \\ 
R$^{2}$ & 0.303 & 0.295 & 0.266 \\ 
Adjusted R$^{2}$ & 0.284 & 0.276 & 0.246 \\ 
\hline \\[-1.8ex] 
\textit{Note:}  & \multicolumn{3}{r}{$^{*}$p$<$0.1; $^{**}$p$<$0.05; $^{***}$p$<$0.01} \\ 
\end{tabular} 
\end{table}

\renewcommand{\arraystretch}{0.95}
\begin{table}[!htbp] \centering 
  \caption{Controlled correlations between different measures on the diversity of visitors and the complexity of amenities} 
  \label{table:si9_table2} 
\begin{tabular}{@{\extracolsep{5pt}}lccc} 
\hline \\[-1.8ex] 
\\[-1.8ex] & Coefficient of variation & Gini & Theil \\ 
\\[-1.8ex] & (1) & (2) & (3)\\ 
\hline \\[-1.8ex] 
 Amenity complexity & 0.069$^{***}$ & 0.035$^{***}$ & 0.023$^{***}$ \\ 
  & (0.016) & (0.007) & (0.005) \\ 
  & & & \\ 
 Centrality of location & 0.116$^{***}$ & 0.051$^{***}$ & 0.029$^{***}$ \\ 
  & (0.011) & (0.005) & (0.003) \\ 
  & & & \\ 
 Nr POIs in category (log) & $-$0.006 & $-$0.004 & $-$0.002 \\ 
  & (0.007) & (0.003) & (0.002) \\ 
  & & & \\ 
 Nr visitors (log) & 0.069$^{***}$ & 0.032$^{***}$ & 0.016$^{***}$ \\ 
  & (0.008) & (0.004) & (0.002) \\ 
  & & & \\ 
 Constant & 0.172$^{***}$ & 0.100$^{***}$ & 0.017$^{*}$ \\ 
  & (0.030) & (0.014) & (0.009) \\ 
  & & & \\ 
\hline \\[-1.8ex] 
Observations & 2,742 & 2,742 & 2,742 \\ 
R$^{2}$ & 0.108 & 0.115 & 0.074 \\ 
Adjusted R$^{2}$ & 0.106 & 0.114 & 0.073 \\ 
\hline \\[-1.8ex] 
\textit{Note:}  & \multicolumn{3}{r}{$^{*}$p$<$0.1; $^{**}$p$<$0.05; $^{***}$p$<$0.01} \\ 
\end{tabular} 
\end{table}

\clearpage
\section*{S10 Diversity of non-local visitors to neighborhoods}

Table~\ref{table:si10_table1} supports the findings presented in Table 1 of the main text, but the models consider only the socio-economic diversity of non-local visitors. This means that we excluded all third place visits to neighborhoods by devices with identified home location inside the same neighborhood. Results are very similar to our main models.

\renewcommand{\arraystretch}{0.95}
\begin{table}[!htbp] \centering 
  \caption{Controlled correlations between the diversity of non-local visitors and the amenity complexity of neighborhoods} 
  \label{table:si10_table1}
\begin{tabular}{@{\extracolsep{5pt}}lccccc} 
\hline \\[-1.8ex] 
\\[-1.8ex] & \multicolumn{5}{c}{Coefficient of variation (non-local visitors only)} \\ 
\\[-1.8ex] & (1) & (2) & (3) & (4) & (5)\\ 
\hline \\[-1.8ex] 
 Neighborhood complexity &  & 0.134$^{***}$ &  &  & 0.135$^{***}$ \\ 
  &  & (0.035) &  &  & (0.036) \\ 
  & & & & & \\ 
 Amenity diversity &  &  & $-$0.169$^{**}$ &  & $-$0.183$^{**}$ \\ 
  &  &  & (0.074) &  & (0.072) \\ 
  & & & & & \\ 
 Avg amenity ubiquity &  &  &  & $-$0.046 & $-$0.013 \\ 
  &  &  &  & (0.048) & (0.049) \\ 
  & & & & & \\ 
 Centrality of location & 0.113$^{***}$ & 0.070$^{**}$ & 0.113$^{***}$ & 0.107$^{***}$ & 0.067$^{**}$ \\ 
  & (0.026) & (0.028) & (0.026) & (0.027) & (0.028) \\ 
  & & & & & \\ 
 Population (log) & $-$0.043$^{*}$ & $-$0.040$^{*}$ & $-$0.038$^{*}$ & $-$0.038$^{*}$ & $-$0.033 \\ 
  & (0.022) & (0.021) & (0.022) & (0.023) & (0.022) \\ 
  & & & & & \\ 
 Nr visitors (log) & 0.039 & 0.031 & 0.047 & 0.037 & 0.039 \\ 
  & (0.029) & (0.028) & (0.029) & (0.029) & (0.028) \\ 
  & & & & & \\ 
 Nr POIs (log) & 0.008 & 0.015 & 0.061$^{*}$ & $-$0.004 & 0.070$^{**}$ \\ 
  & (0.023) & (0.022) & (0.033) & (0.026) & (0.033) \\ 
  & & & & & \\ 
 Constant & 0.403$^{***}$ & 0.337$^{***}$ & 0.330$^{***}$ & 0.436$^{***}$ & 0.267$^{***}$ \\ 
  & (0.055) & (0.056) & (0.063) & (0.065) & (0.073) \\ 
  & & & & & \\ 
\hline \\[-1.8ex] 
Observations & 186 & 186 & 186 & 186 & 186 \\ 
R$^{2}$ & 0.231 & 0.287 & 0.252 & 0.235 & 0.312 \\ 
Adjusted R$^{2}$ & 0.214 & 0.268 & 0.232 & 0.213 & 0.285 \\ 
\hline \\[-1.8ex] 
\textit{Note:}  & \multicolumn{5}{r}{$^{*}$p$<$0.1; $^{**}$p$<$0.05; $^{***}$p$<$0.01} \\ 
\end{tabular} 
\end{table}

\clearpage
\section*{S11 Amenity complexity and the diversity of visitors over 24 months}

\begin{figure}[htb]
\centering
\includegraphics[width=0.95\textwidth]{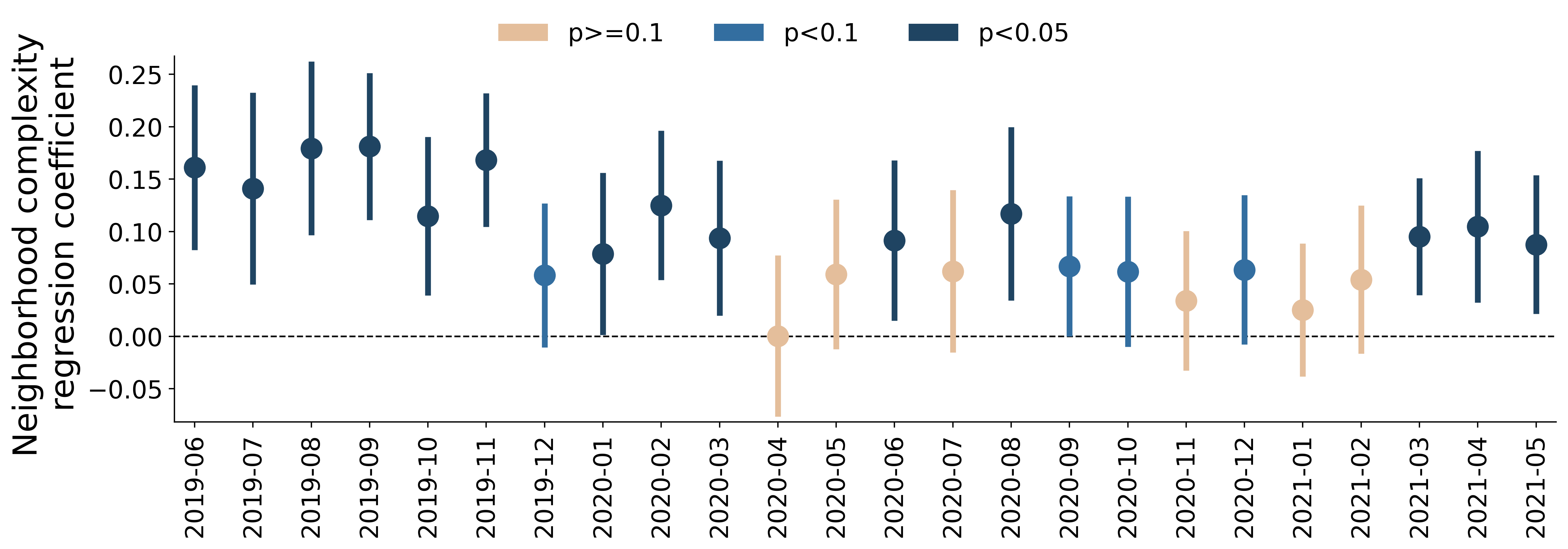}
\caption{Regression coefficient of neighborhood complexity estimated for 24 months by model (2) of Table 1 in the main text. Different colors indicate different levels of significance.}
\label{fig:si11_fig1}
\end{figure}

Figure~\ref{fig:si11_fig1} illustrates the relationship between diversity of visitors to neighborhoods and neighborhood complexity over 24 months. The figure shows the coefficients for neighborhood complexity from the regression setting presented in Table 1 model (2) of the main text. It suggests that neighborhood complexity does not have a significant relationship with the diversity of visitors in 6 out of 24 months.

\begin{figure}[htb]
\centering
\includegraphics[width=0.95\textwidth]{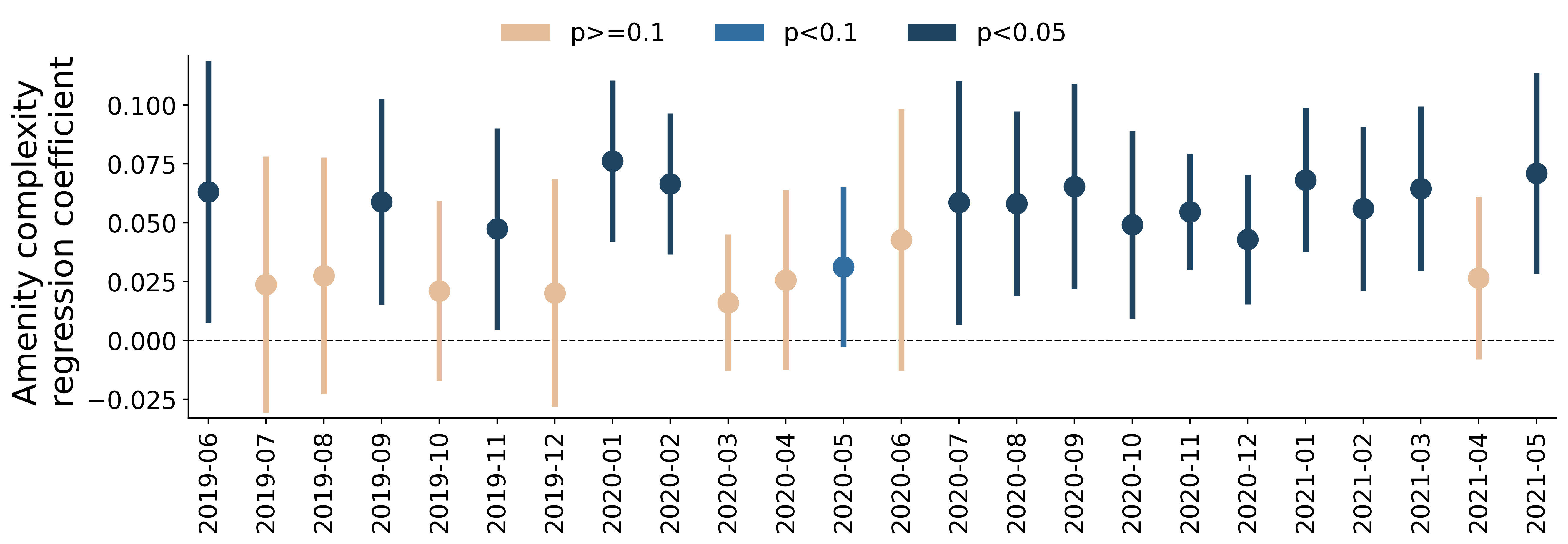}
\caption{Regression coefficient of amenity complexity estimated for 24 months by model (2) of Table 2 in the main text. Different colors indicate different levels of significance.}
\label{fig:si11_fig2}
\end{figure}

Figure~\ref{fig:si11_fig2} illustrates the relationship between diversity of visitors to amenities and amenity complexity over 24 months. The figure represents the coefficients for amenity complexity from the regression setting presented in Table 2 model (2) of the main text. It shows that neighborhood complexity does not have a significant relationship to the diversity of visitors in 8 out of 24 months. The differences between the two figures do not allow us to argue for a strong impact of COVID-19 on visitation patterns.

\clearpage
\section*{S12 Instrumental variable approaches to estimate the relationship \\ between neighborhood complexity and diversity of visitors}

There are a number of potential unobserved factors that influence both neighborhood complexity and the socio-economic diversity of visitors to neighborhoods, creating a potential endogeneity problem when measuring their relationship. To account for this potential endogeneity, we re-estimate our results presented in Table 1 of the main text using different instrumental variable (IV) approaches \citep{angrist1996iv}. The IV estimation approach corrects for possible endogeneity in regression settings by using variables (instruments) that do not belong to the explanatory equation. The requirement  for this is that IVs are uncorrelated to the dependent variable (at least conditionally on the level of the independent variable) but correlated with the main independent variable (in our case, neighborhood complexity).

\textcite{broekel2019complexity} argues that complexity measures constructed from the spatial distribution of activities are potentially endogenous variables in many spatial research settings. At the international level, \textcite{salinas2021distancecomplexity} shows that economic complexity is related to the distance of countries from economic centers, and as a result, complexity indicators show spatial correlation. In the main text we present that our measures of amenity complexity are also connected to the urban centrality of locations.
To account for this endogeneity issue we follow \textcite{stojkoski2022multidimensional}
and introduce an IV which is constructed for each neighborhood by taking the average neighborhood complexity of the three neighborhoods with the most similar specialization pattern to the focal neighborhood. 
The logic behind this IV is that neighborhoods with similar specialization patterns should also have similar neighborhood complexity as the focal neighborhood, but the complexity of the similar neighborhoods should not affect the visitation pattern to the focal neighborhood and should not be connected to other neighborhood specific unobserved characteristics. Table~\ref{table:si12_table1} presents the IV regressions alongside the original OLS regression from the main text. The IV estimation reinforces our original finding, however, the coefficient of neighborhood complexity is less significant.

\renewcommand{\arraystretch}{0.95}
\begin{table}[!htbp] \centering 
  \caption{Diversity of visitors to neighborhoods and neighborhood complexity: instrumental variable approach using similar specialization patterns} 
  \label{table:si12_table1}
\begin{tabular}{@{\extracolsep{5pt}}lccc} 
\hline \\[-1.8ex] 
\\[-1.8ex] & \multicolumn{3}{c}{Coefficient of variation} \\ 
\\[-1.8ex] & \multicolumn{2}{c}{OLS} & IV \\ 
\\[-1.8ex] & (1) & (2) & (3)\\ 
\hline \\[-1.8ex] 
 Neighborhood complexity & 0.125$^{***}$ &  & 0.074$^{*}$ \\ 
  & (0.036) &  & (0.044) \\ 
  & & & \\ 
 Neighborhood complexity (avg of 3 most similar) &  & 0.050$^{*}$ &  \\ 
  &  & (0.026) &  \\ 
  & & & \\ 
 Centrality of location & 0.089$^{***}$ & 0.103$^{***}$ & 0.121$^{***}$ \\ 
  & (0.028) & (0.030) & (0.030) \\ 
  & & & \\ 
 Population (log) & $-$0.050$^{**}$ & $-$0.055$^{**}$ & $-$0.045$^{**}$ \\ 
  & (0.022) & (0.022) & (0.022) \\ 
  & & & \\ 
 Nr visitors (log) & 0.039 & 0.050$^{*}$ & 0.036 \\ 
  & (0.029) & (0.029) & (0.029) \\ 
  & & & \\ 
 Nr POIs (log) & 0.006 & 0.002 & 0.002 \\ 
  & (0.023) & (0.023) & (0.023) \\ 
  & & & \\ 
 Constant & 0.363$^{***}$ & 0.409$^{***}$ & 0.375$^{***}$ \\ 
  & (0.057) & (0.056) & (0.059) \\ 
  & & & \\ 
\hline \\[-1.8ex] 
Observations & 186 & 186 & 186 \\ 
R$^{2}$ & 0.303 & 0.273 & 0.294 \\ 
Adjusted R$^{2}$ & 0.284 & 0.253 & 0.275 \\ 
\hline \\[-1.8ex] 
\textit{Note:}  & \multicolumn{3}{r}{$^{*}$p$<$0.1; $^{**}$p$<$0.05; $^{***}$p$<$0.01} \\ 
\end{tabular} 
\end{table}

In another IV estimation setting, the historical separation of neighborhoods from Budapest is used as an instrument for neighborhood complexity. The current form of Budapest was only realized in 1950, when 59 municipalities gave up their autonomy and were integrated into the capital. We create a dummy variable for each neighborhood and use it as an IV which takes the value of 1 in case the neighborhood only became part of Budapest in 1950 and 0 otherwise. The rationale behind using this variable as an instrument is that municipalities that were independent before 1950 had an amenity structure organized to serve the independent municipality (e.g. independent local administration, marketplaces, libraries or post offices), while the already integrated municipalities could rely more on the central facilities and services of the capital. Consequently, historical separation of neighborhoods could influence their complexity, but not the socio-economic diversity of visitors to neighborhoods today (beyond its indirect effect through complexity). Table~\ref{table:si12_table2} presents the IV regressions using historical separation as an instrument alongside the original OLS regression from the main text. The IV estimation reinforces our original findings, but the coefficient of neighborhood complexity is less significant. Both of our instruments are statistically valid.

\renewcommand{\arraystretch}{0.95}
\begin{table}[!htbp] \centering 
  \caption{Diversity of visitors to neighborhoods and neighborhood complexity: instrumental variable approach using historical separation of neighborhoods} 
  \label{table:si12_table2} 
\begin{tabular}{@{\extracolsep{5pt}}lcc}
\hline \\[-1.8ex] 
\\[-1.8ex] & \multicolumn{2}{c}{Coefficient of variation} \\ 
\\[-1.8ex] & OLS & IV \\
\\[-1.8ex] & (1) & (2)\\ 
\hline \\[-1.8ex] 
 Neighborhood complexity & 0.125$^{***}$ & 0.485$^{**}$ \\ 
  & (0.036) & (0.209) \\ 
  & & \\ 
 Centrality of location & 0.089$^{***}$ & $-$0.004 \\ 
  & (0.028) & (0.072) \\ 
  & & \\ 
 Population (log) & $-$0.050$^{**}$ & $-$0.031 \\ 
  & (0.022) & (0.028) \\ 
  & & \\ 
 Nr visitors (log) & 0.039 & 0.007 \\ 
  & (0.029) & (0.038) \\ 
  & & \\ 
 Nr POIs (log) & 0.006 & 0.024 \\ 
  & (0.023) & (0.030) \\ 
  & & \\ 
 Constant & 0.363$^{***}$ & 0.167 \\ 
  & (0.057) & (0.125) \\ 
  & & \\ 
\hline \\[-1.8ex] 
Observations & 186 & 186 \\ 
R$^{2}$ & 0.303 & $-$0.079 \\ 
Adjusted R$^{2}$ & 0.284 & $-$0.109 \\ 
\hline \\[-1.8ex] 
\textit{Note:}  & \multicolumn{2}{r}{$^{*}$p$<$0.1; $^{**}$p$<$0.05; $^{***}$p$<$0.01} \\ 
\end{tabular} 
\end{table}

\end{document}